\documentclass[12pt]{article}%
\usepackage[nosort]{cite}
\usepackage{graphicx}
\usepackage{multicol}
\usepackage{amsfonts}
\usepackage{amssymb}
\usepackage{amsmath}
\usepackage{heck}
\usepackage{afterpage}
\usepackage{setspace}
\usepackage{verbatim}
\usepackage{color}
\usepackage{longtable}
\usepackage{float}
\usepackage{subcaption}
\usepackage{epsfig}
\usepackage{epstopdf}
\usepackage{adjustbox}
\usepackage{tikz}
\usepackage[margin=1in]{geometry}
\usepackage{titletoc}
\usepackage{hyperref}%
\usepackage{mathrsfs}
\usepackage{dsfont}
\usepackage{algorithm}
\usepackage{algpseudocode}
\usepackage{upgreek}
\usepackage{mathtools}
\usepackage{stmaryrd}
\usepackage{caption}
\providecommand{\U}[1]{\protect\rule{.1in}{.1in}}
\newsavebox{\mysavebox}

\hypersetup{colorlinks,citecolor=black,filecolor=black,linkcolor=black,urlcolor=black}
\usetikzlibrary{decorations.markings}

\numberwithin{equation}{section}

\newcommand{\be}{\begin{equation}}
\newcommand{\ee}{\end{equation}}

\usepackage{tikz}
\usetikzlibrary{arrows}
\usetikzlibrary{arrows.meta}
\usetikzlibrary{arrows,decorations.pathmorphing}
\usetikzlibrary{shapes.geometric,calc,arrows, positioning,shapes.misc,decorations.markings}
\tikzset{
  big arrow/.style={
    decoration={markings,mark=at position 1 with {\arrow[scale=2,#1]{>}}},
    postaction={decorate},
    shorten >=0.4pt},
  big arrow/.default=black}

\pgfdeclarelayer{edgelayer}
\pgfdeclarelayer{nodelayer}
\pgfsetlayers{edgelayer,nodelayer,main}
\tikzstyle{none}=[inner sep=0pt]

\tikzstyle{NodeCross}=[draw, shape=circle, cross out, inner sep=0pt, minimum size=6pt,line width=0.25mm]
\tikzstyle{Circle}=[draw, shape=circle, black, fill=black, inner sep=0pt, minimum size=6pt]
\tikzstyle{circle}=[draw, shape=circle, black, fill=black, inner sep=0pt, minimum size=16pt]
\tikzstyle{Star}=[draw, shape=star, fill=red!30, star points=8, inner sep=0pt, minimum size=8pt]
\tikzstyle{CircleRed}=[draw, shape=circle, black, fill=red, inner sep=0pt, minimum size=6pt]
\tikzstyle{StarP}=[draw={rgb,255: red,128; green,0; blue,128}, shape=star, fill={rgb,256: red,128; green,0; blue,128}, star points=8, inner sep=0pt, minimum size=12pt]
\tikzstyle{ShadedCircRed}=[draw=red, shape=circle, fill={rgb, 255: red,255; green,114; blue, 118}, inner sep=0pt, minimum size=80pt, line width=0.5mm, fill opacity=0.2]
\tikzstyle{ShadedCircRed2}=[draw=red, shape=circle, fill={rgb, 255: red,255; green,114; blue, 118}, inner sep=0pt, minimum size=10pt]
\tikzstyle{ShadedCircRed3}=[draw=black, shape=rectangle, fill={rgb, 255: red,255; green,114; blue, 118}, inner sep=0pt, minimum size=113pt, line width=0.25mm]
\tikzstyle{ShadedCirc}=[draw=red, shape=circle, fill=white, inner sep=0pt, minimum size=45pt,  fill opacity=1.0,  line width=0.5mm]
\tikzstyle{CircleBlue}=[draw, shape=circle, fill=blue, inner sep=0pt, minimum size=6pt]
\tikzstyle{BigCirclePurple}=[draw, shape=circle, fill={rgb,255: red,191; green,0; blue,191}, inner sep=0pt, minimum size=12pt]
\tikzstyle{CirclePurple}=[draw, shape=circle, fill={rgb,255: red,191; green,0; blue,191}, inner sep=0pt, minimum size=5pt]
\tikzstyle{EmptyCircle}=[draw, shape=circle, inner sep=0pt, minimum size=4pt]
\tikzstyle{GreenCircle}=[draw, shape=circle,  fill={rgb,255: red,80; green,200; blue,120}, inner sep=0pt, minimum size=8pt]
\tikzstyle{BrownCircle}=[draw, shape=circle,  fill={rgb,255: red,210; green,105; blue,30}, inner sep=0pt, minimum size=8pt]
\tikzstyle{CirclePurpleSmall}=[draw, shape=circle, fill={rgb,255: red,191; green,0; blue,191}, inner sep=0pt, minimum size=4pt]
\tikzstyle{BigCircleGreen}=[draw, shape=circle, fill=green!30, inner sep=0pt, minimum size=12pt]
\tikzstyle{BigCircleBlue}=[draw, shape=circle, fill=blue!30, inner sep=0pt, minimum size=12pt]
\tikzstyle{BigCircleRed}=[draw, shape=circle, fill=red!30, inner sep=0pt, minimum size=12pt]
\tikzstyle{BigCircleWhite}=[draw, shape=circle, fill=white!30, inner sep=0pt, minimum size=12pt]
\tikzstyle{BrownCircleSmall}=[draw, shape=circle,  fill={rgb,255: red,210; green,105; blue,30}, inner sep=0pt, minimum size=6pt]

\tikzstyle{SmallCircleBrown}=[draw, shape=circle,  fill={rgb,255: red,210; green,105; blue,30}, inner sep=0pt, minimum size=7pt]

\tikzstyle{SmallCircleRed}=[draw, shape=circle, fill=red!30, inner sep=0pt, minimum size=6pt]

\tikzstyle{SmallCircleGreen}=[draw, shape=circle, fill=green!30, inner sep=0pt, minimum size=6pt]

\tikzstyle{DashedLine}=[-, densely dashed, line width=0.25mm]
\tikzstyle{DottedLine}=[-, dotted, line width=0.25mm]
\tikzstyle{ThickLine}=[-, line width=0.25mm]
\tikzstyle{ArrowLineRight}=[-, -{Stealth[scale=1.25]}, line width=0.25mm, scale=5]
\tikzstyle{ArrowLineRed}=[-, draw={rgb,255: red,191; green,0; blue,0}, -{Stealth[scale=1.75]}, line width=0.1mm, scale=5]
\tikzstyle{RedLine}=[-, draw={rgb,255: red,191; green,0; blue,0}, fill=none, line width=0.5mm]
\tikzstyle{DashedLineThin}=[-, densely dashed, line width=0.125mm, fill=none, draw=black]
\tikzstyle{DottedRed}=[-, dotted, draw={rgb,255: red,191; green,0; blue,0}, fill=none, line width=0.25mm]
\tikzstyle{DashedRed}=[-, densely dashed, draw={rgb,255: red,191; green,0; blue,0}, fill=none, line width=0.25mm]
\tikzstyle{BlueLine}=[-, draw={rgb,255: red,0; green,0; blue,191}, fill=none, line width=0.5mm]
\tikzstyle{ArrowLineBlue}=[-, draw={rgb,255: red,0; green,0; blue,191}, -{Stealth[scale=1.75]}, line width=0.1mm, scale=5]
\tikzstyle{GreenDoubleArrow}=[<->, draw={rgb,155: red,0; green,255; blue,0},  line width= 0.5mm, scale=5]
\tikzstyle{RedDoubleArrow}=[<->, draw={rgb,255: red,255; green,0; blue,0},  line width= 0.5mm, scale=5]
\tikzstyle{BlueDottedLight}=[-, dotted, draw={rgb,255: red,0; green,0; blue,191}, fill=none, line width=0.3mm]
\tikzstyle{BrownLine}=[-, draw={rgb,255: red,210; green,105; blue,30}, fill=none, line width=0.5mm]
\tikzstyle{DottedRed}=[-, dotted, draw={rgb,255: red,191; green,0; blue,0}, fill=none, dotted, line width=0.5mm]
\tikzstyle{DottedPurple}=[-, dotted, draw={rgb,255: red,191; green,0; blue,191}, fill=none, dotted, line width=0.5mm]
\tikzstyle{BlueDottedLight}=[-, dotted, draw={rgb,255: red,0; green,0; blue,191}, fill=none, line width=0.5mm]
\tikzstyle{ArrowLinePurple}=[-, draw={rgb,255: red,191; green,0; blue,191}, -{Stealth[scale=1.75]}, line width=0.5mm, scale=5]
\tikzstyle{DashedLineGreen}=[-, densely dashed, draw={rgb,255: red,74; green,103; blue,65}, line width=0.25mm]
\tikzstyle{LineGreen}=[-, draw={rgb,255: red, 74; green,200; blue,65}, line width=0.5mm]
\tikzstyle{ArrowLineGreen}=[-, draw={rgb,255: red,0; green,191; blue,0}, -{Stealth[scale=1.75]}, line width=0.5mm, scale=5]
\tikzstyle{GreenLine}=[-, draw={rgb,255: red,0; green,191; blue,0}, fill=none, line width=0.5mm]
\tikzstyle{PurpleLine}=[-, draw={rgb,255: red,191; green,0; blue,191}, fill=none, line width=0.5mm]
\tikzstyle{PPurpleLine}=[-, draw={rgb,255: red,191; green,0; blue,191}, fill=none, line width=2.5mm]
\tikzstyle{DPurpleLine}=[-, dotted, draw={rgb,255: red,191; green,0; blue,191}, fill=none, line width=0.5mm]
\tikzstyle{SBrownLine}=[-, draw={rgb,255: red,191; green,0; blue,191}, fill=none, opacity=0.35, line width=2.5mm]
\tikzstyle{DottedBlue}=[-, dotted, draw=blue, fill=none, dotted, line width=0.5mm]

\tikzstyle{DashedPurpleLine}=[-, densely dashed, draw={rgb,255: red,191; green,0; blue,191}, fill=none, line width=0.5mm]

\tikzstyle{SmallCircleBlue}=[draw, shape=circle, fill=blue!30, inner sep=0pt, minimum size=5pt]
\tikzstyle{SmallCirclePurple}=[draw, shape=circle, fill={rgb,255: red,191; green,0; blue,191}, inner sep=0pt, minimum size=5pt]

\tikzset{snake it/.style={decorate, decoration=snake}}

\usetikzlibrary{decorations.markings}
\usetikzlibrary{hobby}

\tikzset{
dashstar/.style={
 dash pattern=on 5pt off 5pt,
 postaction={
  decorate,
  decoration={
   markings,
   mark=between positions 9pt and 1 step 10pt with {
     \node[color=red] {*};
   }
  }
 }
},
dashstarstar/.style={ % from marmot's comments
 dash pattern=on 5pt off 10pt,
 postaction={
   decorate,
   decoration={
     markings,
     mark=between positions 10pt and 1
          step 15pt
           with {
            \node at (-2pt,0pt) {\pgfuseplotmark{star}};
            \node at (2pt,0pt) {\pgfuseplotmark{star}};
           }
   }
 }
}
}
\usepackage{pgfplots}
\pgfplotsset{compat=1.16}

\newcommand{\lb}{\left(}
\newcommand{\rb}{\right)}
\newcommand{\lbb}{\left[}
\newcommand{\rbb}{\right]}

\newcommand{\ba}{\begin{aligned}}
\newcommand{\ea}{\end{aligned}}
\newcommand{\Z}{{\mathbb Z}}

\usepackage{algorithmicx, algpseudocode,algorithm}
\usepackage{caption}

\usepackage{float}

\begin{document}

\begin{flushright}
    UUITP-19/25
\end{flushright}

\date{June 2025}

\title{Symmetry Theories, Wigner's Function, \\[4mm] Compactification, and Holography}

\institution{PENN}{\centerline{$^{1}$Department of Physics and Astronomy, University of Pennsylvania, Philadelphia, PA 19104, USA}}
\institution{PENNmath}{\centerline{$^{2}$Department of Mathematics, University of Pennsylvania, Philadelphia, PA 19104, USA}}
\institution{Uppsala}{\centerline{$^{3}$Department of Physics and Astronomy, Uppsala University, Box 516, SE-75120 Uppsala, Sweden}}

\authors{
Jonathan J. Heckman\worksat{\PENN,\PENNmath}\footnote{e-mail: \texttt{jheckman@sas.upenn.edu}},
Max H\"ubner\worksat{\Uppsala}\footnote{e-mail: \texttt{max-elliot.huebner@physics.uu.se}}, and
Chitraang Murdia\worksat{\PENN}\footnote{e-mail: \texttt{murdia@sas.upenn.edu}}
}

\abstract{The global symmetry data of a $D$-dimensional absolute quantum field theory can sometimes
be packaged in terms of a $(D+1)$-dimensional bulk system obtained by extending
along an interval, with a relative QFT$_D$ at one end and suitable
gapped / free boundary conditions at the other end. The partition function
of the QFT$_D$ can then be interpreted as a wavefunction depending on
background fields. However, in some cases, it is not possible or simply cumbersome to fix an absolute form
of the symmetry data. Additionally, it is also of interest to consider entangled and mixed states of relative QFTs as well as entangled and mixed states of gapped / free boundary conditions. We argue that Wigner's quasi-probabilistic function on phase
space provides a physical interpretation of the symmetry data in all such situations. We illustrate
these considerations in the case of string compactifications and holographic systems.}

\maketitle

\enlargethispage{\baselineskip}

\setcounter{tocdepth}{2}

%\tableofcontents

\newpage

\section{Introduction}

Symmetries provide important constraints on a broad range of quantum systems.
One of the recent developments in the subject is the appearance of rich
topological structures connected with the symmetries themselves
\cite{Gaiotto:2014kfa}. This has by now led to vast generalizations of the
standard paradigm of symmetries, which are still being uncovered.

\begin{figure}
\centering
\scalebox{0.8}{
\begin{tikzpicture}
	\begin{pgfonlayer}{nodelayer}
		\node [style=BigCircleBlue] (0) at (-5.25, 0) {};
		\node [style=BigCircleRed] (1) at (-1.75, 0) {};
		\node [style=BigCircleBlue] (2) at (2, 0) {};
		\node [style=BigCircleRed] (3) at (5, 0) {};
		\node [style=none] (6) at (3.5, -2) {(ii)};
		\node [style=none] (7) at (-3.5, -2) {(i)};
		\node [style=none] (8) at (-5.25, 0.625) {$\langle B |$};
		\node [style=none] (9) at (-1.75, 0.625) {$| Z\rangle$};
		\node [style=none] (12) at (5, 0.625) {$|\rho_Z\rangle \!\rangle$};
		\node [style=none] (18) at (-3.5, -0.625) {SymTFT $\mathcal{S}$};
		\node [style=none] (19) at (0, -2.25) {};
		\node [style=none] (20) at (3.35, 0.15) {};
		\node [style=none] (21) at (3.65, -0.15) {};
		\node [style=none] (22) at (5, -0.15) {};
		\node [style=none] (23) at (2, -0.15) {};
		\node [style=none] (24) at (2, 0.15) {};
		\node [style=none] (25) at (5, 0.15) {};
		\node [style=none] (26) at (2, 0.625) {$\langle \!\langle \rho_B|$};
		\node [style=none] (42) at (3.5, -0.625) {$\mathcal{S}\otimes \overline{\mathcal{S}}$};
	\end{pgfonlayer}
	\begin{pgfonlayer}{edgelayer}
		\draw [style=ThickLine] (0) to (1);
		\draw [style=ThickLine] (21.center) to (22.center);
		\draw [style=ThickLine] (20.center) to (24.center);
		\draw [style=ArrowLineRight] (25.center) to (20.center);
		\draw [style=ArrowLineRight] (23.center) to (21.center);
	\end{pgfonlayer}
\end{tikzpicture}
}
\caption{(i): Schematic depiction of the SymTFT formalism and its wavefunction interpretation. We decompress an absolute QFT to a relative QFT  $|Z\rangle$ and a topological / free boundary $\langle B|$ with bulk a SymTFT / SymTh $\mathcal{S}$. The overlap is then $Z(B)$, the partition function with prescribed boundary conditions. (ii): Similarly, mixed states of relative QFTs $\rho_Z=\sum_{ij}z_{ij}|Z_i\rangle \langle Z_j|$  and topological boundary conditions are decompressed into a $|\rho_Z\rangle\!\rangle$ and $\langle \!\langle \rho_B|$ which are ket and bra to the SymTFT / SymTh $\mathcal{S}\otimes \overline{\mathcal{S}}$.}
\label{fig:SymTFTschematic}
\end{figure}
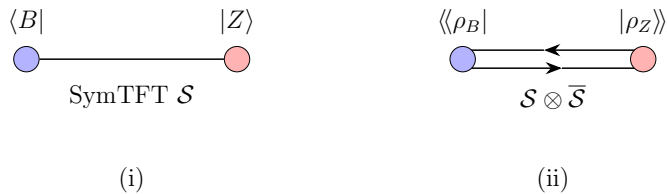

A convenient way to capture many aspects of symmetries is by
working with a higher-dimensional bulk system. For example, for a finite
symmetry $G$ of a $D$-dimensional QFT$_{D}$ on a manifold $M_{D}$, there is a
corresponding symmetry topological field theory SymTFT$_{D+1}$ obtained by
extending along an interval $I\times M_{D}$. At one end, we have a relative QFT$_D$, and at the other, we have a gapped theory. This can be enlarged to cover
situations such as continuous symmetries. One can visualize the two boundaries
as specifying a physical state (i.e., a choice of relative QFT) $\left\vert
Z\right\rangle $ and a gapped / free boundary condition $\left\langle
B\right\vert $. The partition function of the absolute theory is then a wavefunction:
\begin{equation}
Z(B)=\left\langle B|Z\right\rangle .
\end{equation}
There are now many bottom-up and top-down treatments of this basic
framework.\footnote{See e.g., \cite{Reshetikhin:1991tc, TURAEV1992865,
Barrett:1993ab, Witten:1998wy, Fuchs:2002cm, Kirillov2010TuraevViroIA,
Kapustin:2010if,Kitaev2011ModelsFG, Fuchs:2012dt, Freed:2012bs, Heckman:2017uxe, Freed:2018cec,
Gaiotto:2020iye, Apruzzi:2021nmk, Freed:2022qnc, Kaidi:2022cpf, Baume:2023kkf,
Brennan:2024fgj, Heckman:2024oot, Argurio:2024oym, Heckman:2024zdo, Cvetic:2024dzu, Bonetti:2024cjk, Apruzzi:2024htg}\ for a partial list of references to foundational early work, as well as more recent generalizations.}
See subfigure (i) of figure \ref{fig:SymTFTschematic} for a depiction of the SymTFT formalism.

However, in some situations, this formulation can be somewhat awkward. For example,
specifying a particular polarization and thus an absolute QFT$_{D}$ is not
always possible. Examples of this sort include many 6D SCFTs where no
polarization is available.\footnote{Instead, one typically has a vector of
partition functions (see e.g., \cite{Witten:2009at, Tachikawa:2013hya,
DelZotto:2015isa}). Formally speaking, one can always introduce by hand an
auxiliary \textquotedblleft center of mass\textquotedblright\ theory to
produce a single partition function, but this can obscure some physical
features. Adding this center of mass mode is straightforward, for example, for 6D (2,0) SCFTs constructed from M5-branes. For other 6D SCFTs the same construction applies simply because the added center of mass edge mode (which would be added to the topological boundary condition) only couples to the original system via its 2-form symmetry.  }
Another issue is that, especially in
holographic systems, it is often convenient to entertain (at least semi-classically)
an ensemble of theories,\footnote{We remain agnostic as to whether such ensembles are an artifact
of working in the IR, or might simply reflect a necessary approximation in dealing with a highly complex system.
For recent discussions of ensemble averaging in holography from bottom-up and top-down
perspectives, see e.g., \cite{Maldacena:2016hyu, Stanford:2019vob,Saad:2019lba, Balasubramanian:2020lux, Marolf:2020xie, Heckman:2021vzx, Chandra:2022bqq,Schlenker:2022dyo, Benini:2022hzx, Baume:2023kkf}.}
and thus mixed states of relative QFTs and their boundary conditions, e.g.:
\begin{equation}
\widehat{\rho}_{Z}=\underset{i,j}{\sum}z_{ij}\left\vert Z_{i}\right\rangle \left\langle
Z_{j}\right\vert \text{ \ \ and \ \ }\widehat{\rho}_{B}=\underset{i,j}{\sum}%
b_{ij}\left\vert B_{i}\right\rangle \left\langle B_{j}\right\vert .
\end{equation}
In such situations, fixing a single polarization / absolute QFT\ can be cumbersome.

In this note, we address these issues by developing the SymTFT / SymTh formalism
without reference to a fixed polarization.\footnote{SymTh is an abbreviation for Symmetry Theory. By now there are many extra-dimensional constructions associated with the characterization of a system's symmetries, see, e.g., \cite{Baume:2023kkf,
Brennan:2024fgj, Heckman:2024oot, Argurio:2024oym, Heckman:2024zdo, Cvetic:2024dzu, Bonetti:2024cjk, Apruzzi:2024htg} which generalize the original SymTFT discussion. In particular, these bulk theories need not be topological, and we will collectively refer to these constructions as SymThs.} To do this, we formulate
the relevant symmetry data in terms of Wigner's quasi-probabilistic function
\cite{Wigner:1932eb}. This is a function that takes values on the phase space
of a quantum mechanical theory. Integrating over a Lagrangian submanifold of
the phase space then reduces to standard expressions with respect to a fixed
polarization such as $\left\vert Z(B)\right\vert ^{2}$, namely the ``square of the partition function''.\footnote{A perhaps helpful comment is that especially in intrinsically chiral systems it is often more straightforward to define the ``partition function squared'' rather than the partition function itself. Examples include many 2D and 6D conformal field theories.}

The rest of this note is organized as follows. 
In section \ref{sec:2}, we review Wigner's
function in quantum mechanics. In section \ref{sec:3}, we extend this to the
SymTFT / SymTh\ formalism and use it to formulate partition functions for entangled states and ensembles
of relative QFTs and gapped / free boundary conditions. In particular, we use this
setup to study entanglement of SymTFTs / SymThs without reference to a fixed polarization.
In section \ref{sec:4}, we examine two examples motivated by higher-dimensional gravity -- string compactification and holography.

\section{Wigner's Quasi-Probabilistic Distribution}
\label{sec:2}

We first review Wigner's quasi-probabilistic
distribution \cite{Wigner:1932eb}. For our purposes, the main utility of this formalism is that it allows us to
visualize a distribution function on all of phase space, namely,
without fixing a particular polarization (i.e., a position or momentum basis
for states). This will be useful in the context of the SymTFT formalism since
then we need not specify a particular absolute quantum field theory.

\subsection{Continuum Case}

We begin by reviewing some general aspects of Wigner's quasi-probabilistic function in the continuum \cite{Wigner:1932eb}.
The main ideas can already be conveyed in quantum mechanics.
Introduce position and momentum operators $\widehat{q}$ and
$\widehat{p}$ with commutation relation $\lbb\widehat{q},\widehat{p
}\kern0.13em \rbb=i$. We label the eigenstates for these operators as $\left\vert
q\right\rangle $ and $\left\vert p\right\rangle $. For an arbitrary state, with
density matrix
\begin{equation}
\widehat{\rho}=
{\displaystyle\sum\limits_{i}}
\,\rho_{i}\left\vert i\right\rangle \left\langle i\right\vert ,
\end{equation}
we introduce the quasi-probability function:
\begin{equation}
\label{eq:Wquasifunc}
W_{\rho}(q,p)=\frac{1}{\pi}\int dq^{\prime}\text{ }\left\langle
q-q^{\prime}\right\vert \widehat{\rho}\left\vert q
+q^{\prime}\right\rangle e^{2ipq^{\prime}}.
\end{equation}
We can obtain the matrix elements for the density matrix by integrating against the position and momentum, respectively:
\begin{align}\label{eq:Key1}
\int dp\text{ }W_{\rho}(q,p) &  =\left\langle q\right\vert
\widehat{\rho}\left\vert q\right\rangle , \\\label{eq:Key2}
\int dq\text{ }W_{\rho}(q,p) &  =\left\langle p\right\vert
\widehat{\rho}\left\vert p\right\rangle .
\end{align}
Given a classical function $g(q,p)$, we produce a corresponding
Weyl-ordered operator $\widehat{G}$. The expectation value of $\widehat{G}$ is given by the integral over all of the phase space.%
\begin{equation}
\text{Tr}(\widehat{\rho}\:\!\widehat{G})=\int dq dp\text{ }W_{\rho
}(q,p)g(q,p).\label{eq:greatexpectations}%
\end{equation}

We now give a slightly more general presentation of $W_{\rho}(q,p)$. To begin, we introduce the translation operators:
\begin{align}\label{eq:UV}
U_{q}  &  =\exp(-iq\:\!\widehat{p}\kern0.13em )\text{ \ \ with \ \ }U_{q
}\left\vert q^{\prime}\right\rangle =\left\vert q+q^{\prime
}\right\rangle , \\
V_{p}  &  =\exp(+ip\:\!\widehat{q}\kern0.13em )\text{ \ \ with \ \ }V_{p}\left\vert
p^{\prime}\right\rangle =\left\vert p+p^{\prime}\right\rangle ,
\end{align}
as well as the charge conjugation operator $\mathcal{C}$ that acts as:%
\begin{equation}\label{eq:Cop}
\mathcal{C}\left\vert q\right\rangle =\left\vert -q\right\rangle
\text{ \ \ and \ \ }\mathcal{C}\left\vert p\right\rangle =\left\vert
-p\right\rangle \text{.}%
\end{equation}
Then, we can write $W_{\rho}(q,p)$ as an expectation value of operators:
\begin{align}\label{eq:operatorspulledout}
W_{\rho}(q,p)  &  =\frac{1}{\pi}\int dq^{\prime}\text{ }\left\langle
q-q^{\prime}\right\vert \widehat{\rho}\left\vert q
+q^{\prime}\right\rangle e^{2ip q^{\prime}}\\
&  =\frac{1}{\pi}\int dq^{\prime}\text{ }\left\langle -q^{\prime}\right\vert
V_{p}^{\dag}U_{q}^{\dag}\;\!\widehat{\rho}\,U_{q}V_{p}\left\vert
q^{\prime}\right\rangle .
\end{align}
Now, since $\left\langle q^{\prime}\right\vert \mathcal{C} = \left\langle
-q^{\prime}\right\vert $, we also have:
\begin{equation}
W_{\rho}(q,p)  = \frac{1}{\pi}\int dq^{\prime}\text{ }\left\langle
q^{\prime}\right\vert \mathcal{C}V_{p}^{\dag}U_{q}^{\dag}\,\widehat{\rho
}\,U_{q}V_{p}\left\vert q^{\prime}\right\rangle ,
\end{equation}
or:%
\begin{equation}
W_{\rho}(q,p)  = \frac{1}{\pi} \text{Tr}(\widehat{\rho}\,U_{q}V_{p}%
\;\!\mathcal{C}V_{p}^{\dag}U_{q}^{\dag}). \label{eq:traceexpression}
\end{equation}

An interesting feature of this expression is that it allows us to treat the
operators $U$ and $V$ on the same footing. In the context of partition
functions and SymTFT\ expressions, this is helpful because one often has to
specify a polarization. This is especially awkward in systems where an
absolute polarization may not even be available.

Along these lines, it is helpful to repackage our expression slightly
differently. Introduce the \textquotedblleft doubled
operator\textquotedblright\ which ranges over all of the phase space:%
\begin{equation}\label{eq:doubledop}
\mathbb{U}_{q,p}=\exp(-iq\:\!\widehat{p}+ip\:\!\widehat{q}\kern0.13em).
\end{equation}
Then, using Baker-Campbell-Hausdorff formula, we have:%
\begin{equation}\label{eq:doubledop2}
U_{q}V_{p}=\mathbb{U}_{q,p}\exp\left(  -\frac{i}{2}q
p\right) ,
\end{equation}
so it follows that:%
\begin{equation}
\label{eq:WignerWithDoubled}
W_{\rho}(q,p)=\frac{1}{\pi}\text{Tr}(\widehat{\rho}\,\mathbb{U}_{q,p}\;\!\mathcal{C}\;\!%
\mathbb{U}_{q,p}^{\dag}),
\end{equation}
because the c-number phase cancels out.

This last formulation is particularly suggestive. Let $\Phi^{k} = (q,p)$
denote a coordinate on the classical phase space with symplectic pairing:\footnote{The following expression clearly generalizes to the case
where we have a higher-dimensional phase space.}
\begin{equation}
\mu_{kl}\Phi_{(a)}^{k}\Phi_{(b)}^{l}\equiv q_{(a)}p_{(b)}-p
_{(a)}q_{(b)}.
\end{equation}
Then, we have:%
\begin{equation}
\label{eq:Important}
 \mathbb{U}_{\Phi}\equiv \mathbb{U}_{q,p}=\exp\left(  -i\mu_{kl}\Phi
^{k}\widehat{\Phi}^{l}\right)  ,
\end{equation}
where $\widehat{\Phi} \equiv (\widehat{q}, \widehat{p}\:\!)$. Hence:
\begin{equation}
W_{\rho}(\Phi)=\frac{1}{\pi}\text{Tr}(\widehat{\rho}\,\mathbb{U}_{\Phi}\:\!\mathcal{C}\:\!%
\mathbb{U}_{\Phi}^{\dag}). \label{eq:WignerMixed}%
\end{equation}
Specifying a Lagrangian submanifold $L$ of the full phase space, integrating
over $L$ results in a standard probability distribution / expectation value on the conjugate variables.

\subsection{Generalizations}

In many cases, the observables of interest are actually discretized. This leads to some technical complications compared with the
discussion given above. Nevertheless, one can retain much of the same formal
structure, albeit at the expense of working with a somewhat more general class
of Wigner functions. Here, we discuss relevant generalizations in the continuum case.

Indeed, the primary requirement we have for any quasi-probabilistic function on phase space is that we correctly produce the correct quantum mechanical
expectation values upon integrating / summing over the appropriate set of
values. Along these lines, one can introduce a generalized family of Wigner
functions labeled by pairs $r,s\in\mathbb{R}$ given explicitly
by:\footnote{For further discussion on generalizations and extensions to
Wigner's quasi-probability function see references \cite{WOOTTERS19871, Bouzouina1996,
Bianucci2001DiscreteWF}. The necessary non-uniqueness of Wigner's function, as quantified by $r$ and $s$, was already noted by Wigner in his original work \cite{Wigner:1932eb}.}
\begin{equation}
W_{\rho}^{[r,s]}(q,p)=\frac{r+s}{2\pi}\int dq^{\prime}\left\langle
q-rq^{\prime}\right\vert \widehat{\rho}\left\vert q+sq^{\prime}\right\rangle
e^{(r+s)ipq^{\prime}}.
\end{equation}
In \eqref{eq:Wquasifunc} we have $(r,s)=(1,1)$.

This expression can also be recast in terms of an expectation value over
suitable symmetry operators. The only difference is that we also have to introduce a
rescaling operator $\Delta_{\lambda}$ with $\lambda\in\mathbb{R}$ that acts
on the states as:%
\begin{equation}
\Delta_{\lambda}\left\vert q\right\rangle =\left\vert \lambda q\right\rangle .
\end{equation}
Proceeding through a similar set of steps to those already outlined in the preceding subsection, we now have:%
\begin{equation}
\label{eq:GeneralWignerFunction}
W_{\rho}^{[r,s]}(q,p)=\frac{r+s}{2\pi}\text{Tr}(\mathcal{C}\:\!\Delta_{r}^\dagger V_{p}^\dagger U_{q}^\dagger %
\widehat{\rho}\;\!U_{q}V_{p}\Delta_{s}).
\end{equation}
Note that $\Delta_1=\text{Id}$, the identity operator, and $\Delta_{-1}=\mathcal{C}$, the charge conjugation operator.

Following analogous steps that produced \eqref{eq:WignerMixed}, we can favorably rewrite this expression further. Introduce the operator
\begin{equation}
\mathbb{U}_{\Phi}^{[t]} \equiv \mathbb{U}_\Phi\Delta_t\,,
\end{equation}
we have:
\begin{equation}
\label{eq:WignerFunctionPres1}
W_{\rho}^{[r,s]}(\Phi)=\frac{r+s}{2\pi}\text{\:\!Tr}\!\lb \widehat{\rho}\,\mathbb{U}_{\Phi}^{[s]} \:\! \mathcal{C} \:\! \mathbb{U}_{\Phi}^{[r] \, \dagger}\rb.
\end{equation}

\subsection{Discretized Case}

%In many situations of interest one is interested in discretized field configurations.
We now explain how to set up the formalism of Wigner's function in discrete setups. Instead of working directly with position and momentum variables, it is more natural to work with the exponentiated operators:\footnote{Roughly speaking, the commutation relations for $\widehat{q}$ and $\widehat{p}$ are assumed to take the form $[\widehat{q}, \widehat{p}\;\!] = 2 \pi i  / N$. Of course, these are to be viewed as ``mod $N$'' expressions, so it is really more appropriate to work in terms of their exponentiated analogs.}
\begin{equation}
U_{e} = \exp{\left(i e \widehat{p}\;\! \right)} \,\,\, \text{and} \,\,\, V_{m} = \exp{\left(-i m \widehat{q}\;\! \right)},
\end{equation}
with:
\begin{equation}
 V_{m} U_{e} V_{m}^{-1} U_{e}^{-1} = \exp{\left( \frac{2 \pi i}{N} e m \right)},
\end{equation}
where the position and momenta eigenstates $\vert e \rangle$ and $\vert m \rangle$ naturally take values over the integers mod $N$. These vectors span the phase space $\Z_N\times \Z_N$. Compared with the continuum case, the only changes are that now we have finite sums, and the inner product between a position and a momentum eigenstate is:
\begin{equation}
\langle e \vert m \rangle = \frac{1}{\sqrt{N}} \exp\left( \frac{2 \pi i }{N} em \right).
\end{equation}
Other than this, all of the manipulations from the continuum case carry over essentially unchanged.
The only complication is that we must work with $W_{\rho}^{[r,s]}(e,m)$ with $r + s$ relatively prime to $N$, otherwise inversion of Fourier transforms breaks down. Further, given such a good pair, there is a change in normalization\footnote{Simply because for all units of $g\in \Z_N$, as characterized by $\gcd(g,N)=1$, we have:
\begin{equation}
\frac{1}{N}\sum_{n=0}^{N-1}e^{\frac{2\pi i}{N} gn(n'-n'')}=\delta_{n',n''}\,.
\end{equation}} compared to \eqref{eq:WignerFunctionPres1}, we have
\begin{equation}
\label{eq:WignerFunctionPres2}
W_{\rho}^{[r,s]}(\Phi)=\frac{1}{N}\text{\:\!Tr}\!\lb \widehat{\rho}\,\mathbb{U}_{\Phi}^{[s]} \:\! \mathcal{C} \:\! \mathbb{U}_{\Phi}^{[r] \, \dagger}\rb\,,
\end{equation}
in the discrete case.

\subsection{Wavefunction Interpretation for Wigner's Functions}
\label{ssec:WignerAsWave}

We now recast Wigner's function as a wavefunction. This reinterpretation of Wigner's function will later serve in SymTFT / SymTh discussion in making the corresponding symmetry sandwich manifest, including its bulk topological theory, and also specify a natural basis to expand boundary conditions in. In preparation for that discussion, we concentrate on the discrete case here. The continuum case can be treated similarly.

Consider a discretized setup with $N$-dimensional Hilbert space $\mathcal{H}$ and the mixed state $\widehat{\rho}=\sum_i\rho_i|i\rangle \langle i|$. Then we can consider the doubled Hilbert space  $\mathcal{H}\otimes \overline{\mathcal{H}}$ where the overline indicates CPT or CRT relevant to $\mathcal{C}$. We define the pure states:\footnote{We comment that the pure states defined below are not the purifications of mixed states in the original Hilbert space. For example, the purification would have been $\Sigma_i \sqrt{\rho_i} \:\! |\:\!i\:\!\rangle \otimes |\:\!\overline{i}\:\! \rangle\ $. Rather, we have used the operator state map \cite{preskill2015lecture}, resulting in the so-called Choi-state \cite{CHOI1975285}.
}
\begin{equation} \label{eq:purification}
\begin{aligned}
|\rho  \rangle\!\rangle &\equiv \mathcal{N}_{\rho} \sum_i\rho_i\:\! |\:\!i\:\!\rangle \otimes |\:\!\overline{i}\:\! \rangle\,, \\
 \langle \!\langle \Phi^{[r,s]}|  &\equiv \mathcal{N}_{\Phi^{[r,s]}}\sum_{e'} \langle {e}^{\;\!\prime}|\:\!\mathcal{C}\:\!\mathbb{U}_{\Phi}^{[r] \, \dagger}  \otimes \langle e'| \mathbb{U}_{\Phi}^{[s] \, \dagger} \,,
\end{aligned}
\end{equation}
where we have introduced two normalization factors $\mathcal{N}_{\rho}$ and $\mathcal{N}_{\Phi^{[r,s]}}$. We fix $\mathcal{N}_{\Phi^{[r,s]}}$ by requiring orthonormality on phase space:
\begin{equation}
 \langle \!\langle \Phi_1^{[r,s]}|  \Phi_2^{[r,s]} \rangle \!\rangle=\delta_{\Phi_1,\Phi_2}\,.
\end{equation}

Next, note that the bra $ \langle \!\langle \Phi^{[r,s]}| $ depends, after fixing the pair $(r,s)$, only on a point $\Phi$ of the original phase space. In contrast, the original $\langle e |$ depends on the existence of a Lagrangian subset of phase space (which distinguishes position and momentum states) and $e$ then labels a point within that Lagrangian subset.

The bra-ket notation for $|\cdot \rangle\!\rangle$ and $|\cdot \rangle$ distinguishes between vectors of the doubled Hilbert space $\mathcal{H}\otimes \overline{\mathcal{H}}$ and the original Hilbert space $\mathcal{H}$, respectively. These two states contract to give the previously considered Wigner's function with respect to the mixed state $\rho$:
\begin{equation}
 \langle \!\langle \Phi^{[r,s]}| \rho\rangle\!\rangle  = W_{\rho}^{[r,s]}(\Phi)\,.
\end{equation}
Here definitions are such that $\langle e|\:\!O\:\!|\:\!\overline i\:\!\rangle= \langle  i|O^\dagger |e\rangle$ for operators $O$. For example, we have $|\bar e\rangle =|e\rangle$ for basis vectors and complex conjugation according to $\langle e|\:\!\overline i\:\!\rangle=\langle i|e \rangle= \overline{\langle e|i \rangle} $.

Given a classical phase space function $f(\Phi)$, we can now naturally associate it to the covector
\begin{equation}
 \langle \!\langle f^{[r,s]}| \equiv \sum_{\Phi} f(\Phi)  \langle \!\langle \Phi^{[r,s]}|
\end{equation}
with the sum over all of phase space. When discussing SymTFTs / SymThs this will allow us to map functions on their phase space to mixed state boundary conditions. Here, we simply note that we have
\begin{equation}
 \langle \!\langle f^{[r,s]}|  \rho\rangle\!\rangle=\sum_\Phi f(\Phi)  \langle \!\langle \Phi^{[r,s]}|  \rho\rangle\!\rangle=\sum_\Phi f(\Phi) W_{\rho}^{[r,s]}(\Phi)\,.
\end{equation}

\subsection{Illustrative Example: Abelian Chern-Simons Theory}
\label{sec:ABCS}

We now discuss abelian Chern-Simons theory as an illustrative example. The quantization of this theory can be somewhat subtle since there is no
single ``split'' of the phase space into position and momentum variables. See in particular \cite{Witten:1988hf, Belov:2005ze, Kapustin:2010hk} for a careful treatment. We also comment that these considerations naturally extend to theories with a $(2k+1)$-form potential and a Chern-Simons-like action, as in references \cite{Belov:2006jd, Belov:2006xj, Monnier:2013kna, Heckman:2017uxe, Apruzzi:2022dlm}. In particular, the case of a three-form potential captures symmetry data in 6D superconformal field theories (see \cite{DelZotto:2015isa, Heckman:2017uxe, Apruzzi:2022dlm}).

Consider the level $N$ action for a Spin-Chern-Simons theory:\footnote{These considerations generalize to $2k+1$-form potentials and their Chern-Simons-like actions. In this case, one demands a $\mathsf{Wu}$ structure (rather than $\mathsf{Spin}$ structure) on the manifold $M_{4k+3}$ which is viewed as the boundary of a manifold $M_{4k+4}$.}
\begin{equation}
S_{\text{3D}}=\frac{N}{4\pi} \int_{M_3} A dA.
\end{equation}
We consider the spacetime $M_3=\mathbb{R}_\tau\times\Sigma$ with genus $g$ Riemann surface $\Sigma$ and time coordinate $\tau$.

Next, we spell out the different quantities introduced in the earlier sections. After imposing the Gauss' law constraint, we canonically quantize the variables. This results in the conjugate variables $\widehat{q}= \widehat{a}_x$ and $\widehat{p} = \widehat{a}_y$ and the equal time commutator:
\begin{equation}\label{eq:ptAlg}
\big[ \widehat{a}_x(z_1),\widehat{a}_y(z_2)\big]=\frac{2 \pi i}{N} \delta^2(z_1-z_2)\,,
\end{equation}
where locally $A|_{\Sigma}=a_xdx+a_ydy$ on some spatial patch of $\Sigma$ with local coordinates $z=(x,y)$. The algebra implied by \eqref{eq:ptAlg} for gauge-invariant Wilson loops is readily realized on a Hilbert space $\mathcal{H}(\Sigma)$ of complex dimension
\begin{equation}
\text{dim}_{\;\!\mathbb{C}} \mathcal{H}(\Sigma)=N^g\,.
\end{equation}
The basis for this Hilbert space can be given explicitly using the symplectic basis of $H_1(\Sigma)$, as generated by A-cycles $\alpha_i$ and B-cycles $\beta_j$, which can be arranged to intersect such that
\begin{equation}
\alpha_i\cdot \beta_j=\epsilon_{ij}\,, \qquad \alpha_i\cdot \alpha_j=0\,,  \qquad \beta_i \cdot \beta_j=0\,,
\end{equation}
where $i,j=1,\dots, g$ and $H_1(\Sigma)\cong \Z^{2g}$. Basis vectors of $\mathcal{H}(\Sigma)$ are then further characterized by quantized holonomies about the A-cycles or B-cycles (or some mixture of the two).

We denote an orthonormal electric basis by $|i,e_i \rangle$ where $i=1,\dots, g$ runs over the A-cycles and $n_i=0,\dots,N-1$ labels the holonomy along the $i$-th A-cycle $\alpha_i$ which is an $N$-th root of unity. We denote an orthonormal magnetic basis by $|j,m_j \rangle$ where $j=1,\dots, g$ runs over the B-cycles $\beta_j$ and $n_k=0,\dots,N-1$ labels the holonomy along the $j$-th B-cycle, which is also an $N$-th root of unity. Then we have the inner product
\begin{equation}
\langle i,e_i| j,m_j \rangle=\frac{1}{\sqrt{N}}\exp\lb \frac{2\pi i}{N} \delta_{ij} e_im_j\rb\,.
\end{equation}

The conjugate variables take values in the phase space $\Z_{N}^g\times \Z_{N}^g$. Explicitly, we have the parametrization $e=(e_1,\dots, e_g)$ and $m=(m_1,\dots,m_g)$ with $e_i,m_j\in \Z_N$. Given the density matrix $\widehat{\rho}$, we can now construct Wigner's functions. For simplicity, consider the example of the torus $\Sigma=T^2$ for which we have the phase space $\Z_N\times \Z_N$ with the Wigner functions
\begin{equation}
W_{\rho}^{[r,s]}(e,m)= \frac{1}{N} \sum_{e^{\prime}=0}^{N-1} \left\langle
e-re^{\prime}\right\vert \widehat{\rho}\left\vert
e+se^{\prime}\right\rangle
e^{\frac{2\pi i}{N}(r+s)me^{\prime}}\,.
\end{equation}
Observe that given $N$ we must restrict the pair $(r,s)$ to satisfy $\text{gcd}(N,r+s)=1$ to obtain a good Wigner's function as otherwise the analogs of \eqref{eq:Key1}, \eqref{eq:Key2} and \eqref{eq:greatexpectations} receive additional contributions to their righthand sides. For example, whenever $N$ is odd, we can continue to use $r,s = 1$ (the standard presentation). However, when $N$ is even, it is perhaps most minimal to use $r = 1$ and $s = 0$, but this is only a mild inconvenience.

\section{Wigner's Function and SymTFTs / SymThs}
\label{sec:3}

Our discussion so far has focused on some general features of Wigner's quasi-probability distribution in quantum mechanics. We now apply it to the SymTFT / SymTh formalism of QFTs.

The basic object of interest in this context is the partition function of a
$D$-dimensional QFT$_{D}$. For some global symmetries $G$ of an absolute
quantum field theory, we can introduce background values $e$ for the global
symmetries in an electric polarization of the theory. Then, we can evaluate
the partition function with respect to this choice of background values $Z(e)$.
This can, in turn, be interpreted as the evaluation of a path integral for a
gapped / free theory in the bulk with suitable electric\ boundary conditions
dictating the overlap of states:
\begin{equation}
Z(e)=\left\langle e|Z\right\rangle .
\end{equation}
We refer to $\left\vert Z\right\rangle $ as the physical boundary condition (namely a relative quantum field theory) and $\left\langle e\right\vert $ as the gapped / free boundary condition.
Instead of working with respect to this fixed polarization, we can instead
work in a different one by gauging a non-anomalous symmetry. In the language
of quantum mechanics, this gauging just involves summing over the choices of
$e$ to reach:
\begin{equation}
Z(m)=\underset{e}{\sum}\left\langle m|e\right\rangle \left\langle
e|Z\right\rangle ,
\end{equation}
i.e., we have a change of polarization.\footnote{Sometimes this gauging
procedure is obstructed in the sense that the absolute QFT may possess an
anomaly. This is not much of a concern if we broaden our scope to simply
consider a $(D+1)$-dimensional bulk system with the relative QFT$_{D}$ now
viewed as an edge mode. In this broader setting, we simply have the mechanism
of anomaly inflow. The anomaly then tells us that we cannot always compress
the SymTFT$_{D+1}$\ interval.} Considering the pure state $\widehat{\rho}=\left\vert
Z\right\rangle \left\langle Z\right\vert $ in situations with an absolute polarization we have
\begin{equation}
\label{eq:Sum}
\sum_{m}W_Z(e,m)=| Z(e)|^2\,, \qquad \sum_{e}W_Z(e,m)=| Z(m)|^2\,. \qquad
\end{equation}
In what follows, we shall freely interchange
between integration / summation notation when the context and local
conventions are clear.

However, sometimes we cannot work with respect to an absolute quantum field theory.
An example of this sort are most 6D SCFTs (see e.g., \cite{Heckman:2018jxk,
Argyres:2022mnu} for reviews). In these cases, one typically has a vector of
partition functions.\footnote{In some cases, however, one can still fix an absolute
theory. See \cite{Lawrie:2023tdz} for recent discussion on this point.} An alternative is to introduce an auxiliary
\textquotedblleft center of mass\textquotedblright\ degree of freedom so as to
have a single partition function. In any case, the presence of a non-trivial
defect group \cite{DelZotto:2015isa} without a fixed notion of polarization
makes the formulation in terms of an absolute quantum field theory somewhat
more awkward. That being said, there is clearly a well-posed notion of the
corresponding relative quantum field theories, as well as a notion of defects
and symmetry operators, both from a bottom-up and top-down perspective.\footnote{See e.g., \cite{DelZotto:2015isa, Heckman:2022suy, Heckman:2022muc, Lawrie:2023tdz, Bonetti:2024etn}.}

\begin{figure}
\centering
\scalebox{0.8}{
\begin{tikzpicture}
	\begin{pgfonlayer}{nodelayer}
		\node [style=none] (0) at (-5, -1.75) {};
		\node [style=none] (1) at (-5, 0.25) {};
		\node [style=none] (2) at (-4, 1.75) {};
		\node [style=none] (3) at (-4, -0.25) {};
		\node [style=none] (4) at (4, -1.75) {};
		\node [style=none] (5) at (4, 0.25) {};
		\node [style=none] (6) at (5, 1.75) {};
		\node [style=none] (7) at (5, -0.25) {};
		\node [style=none] (8) at (-0.5, -1.75) {};
		\node [style=none] (9) at (-0.5, 0.25) {};
		\node [style=none] (10) at (0.5, -0.25) {};
		\node [style=none] (11) at (0.5, 1.75) {};
		\node [style=none] (12) at (0.75, -1.75) {};
		\node [style=none] (13) at (0.75, 0.25) {};
		\node [style=none] (14) at (1.75, -0.25) {};
		\node [style=none] (15) at (1.75, 1.75) {};
		\node [style=none] (16) at (-1.75, -1.75) {};
		\node [style=none] (17) at (-1.75, 0.25) {};
		\node [style=none] (18) at (-0.75, -0.25) {};
		\node [style=none] (19) at (-0.75, 1.75) {};
		\node [style=none] (20) at (5.75, 0) {$|Z\rangle$};
		\node [style=none] (21) at (-5.75, 0) {$\langle Z| $};
		\node [style=none] (22) at (0.5, 2.5) {$\mathcal{C}$};
		\node [style=none] (23) at (1.75, 2.5) {$ \mathbb{U}_{\Phi}^\dagger$};
		\node [style=none] (24) at (-0.75, 2.5) {$ \mathbb{U}_{\Phi}$};
		\node [style=none] (25) at (0, -2.25) {};
		\node [style=BigCircleRed] (26) at (-4.5, -4) {};
		\node [style=BigCircleRed] (27) at (4.5, -4) {};
		\node [style=BigCircleGreen] (28) at (-1.25, -4) {};
		\node [style=BigCircleGreen] (29) at (0, -4) {};
		\node [style=BigCircleGreen] (30) at (1.25, -4) {};
		\node [style=none] (31) at (5.25, -4) {$|Z\rangle$};
		\node [style=none] (32) at (-5.25, -4) {$\langle Z| $};
		\node [style=none] (33) at (0, -3.375) {$\mathcal{C}$};
		\node [style=none] (34) at (1.25, -3.375) {$ \mathbb{U}_{\Phi}^\dagger$};
		\node [style=none] (35) at (-1.25, -3.375) {$ \mathbb{U}_{\Phi}$};
		\node [style=none] (36) at (-7.75, 0) {(i)};
		\node [style=none] (37) at (-7.75, -4) {(ii)};
		\node [style=none] (38) at (0, -4.75) {};
	\end{pgfonlayer}
	\begin{pgfonlayer}{edgelayer}
		\filldraw[fill=red!30, draw=red!30]  (-5, -1.75) -- (-5, 0.25) -- (-4, 1.75) -- (-4, -0.25) -- cycle;
		\filldraw[fill=green!30, draw=green!30]  (-1.75, -1.75) -- (-1.75, 0.25) -- (-0.75, 1.75) -- (-0.75, -0.25) -- cycle;
		\filldraw[fill=green!30, draw=green!30]  (-.5, -1.75) -- (-0.5, 0.25) -- (0.5, 1.75) -- (0.5, -0.25) -- cycle;
		\filldraw[fill=green!30, draw=green!30]  (0.75, -1.75) -- (0.75, 0.25) -- (1.75, 1.75) -- (1.75, -0.25) -- cycle;
		\filldraw[fill=red!30, draw=red!30]  (4, -1.75) -- (4, 0.25) -- (5, 1.75) -- (5, -0.25) -- cycle;
		\draw [style=ThickLine] (0.center) to (4.center);
		\draw [style=ThickLine] (4.center) to (5.center);
		\draw [style=ThickLine] (5.center) to (1.center);
		\draw [style=ThickLine] (1.center) to (2.center);
		\draw [style=ThickLine] (2.center) to (6.center);
		\draw [style=ThickLine] (6.center) to (5.center);
		\draw [style=ThickLine] (6.center) to (7.center);
		\draw [style=ThickLine] (7.center) to (4.center);
		\draw [style=ThickLine] (1.center) to (0.center);
		\draw [style=DottedLine] (0.center) to (3.center);
		\draw [style=DottedLine] (3.center) to (2.center);
		\draw [style=DottedLine] (3.center) to (7.center);
		\draw (9.center) to (11.center);
		\draw (11.center) to (10.center);
		\draw (10.center) to (8.center);
		\draw (8.center) to (9.center);
		\draw (13.center) to (15.center);
		\draw (15.center) to (14.center);
		\draw (14.center) to (12.center);
		\draw (12.center) to (13.center);
				\draw (17.center) to (19.center);
		\draw (19.center) to (18.center);
		\draw (18.center) to (16.center);
		\draw (16.center) to (17.center);
		\draw [style=ThickLine] (26) to (27);
	\end{pgfonlayer}
\end{tikzpicture}
}
\caption{Wigner's function for the pure state $\widehat{\rho}=|Z\rangle \langle Z|$ avoids any reference to topological / free boundary conditions and can be rephrased purely in reference to two sets of physical boundary conditions (red). The operators $ \mathbb{U}_{\Phi}, \mathcal{C},  \mathbb{U}_{\Phi}^\dagger $ may be interpreted as $(-1)$-form symmetry operators which are codimension-one interfaces (green) in the SymTFT / SymTh slab. Subfigures (i) and (ii) show the same image, with the latter sketched in fewer dimensions. }
\label{eq:AvoidPolarization}
\end{figure}
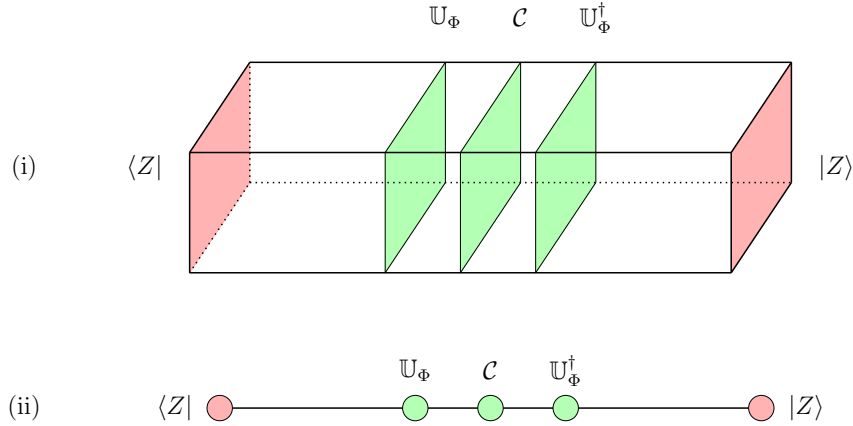

With this in mind, we now observe that so long as we have a field
configuration space, we can still speak of Wigner's function and the associated basis of states $\langle \!\langle \Phi^{[r,s]}|$ of \eqref{eq:purification}, and therefore speak, for example,
of a corresponding square of a partition function.

More precisely for some mixed state $\widehat{\rho}_Z=\sum_{ij}z_{ij}|Z_i\rangle \langle Z_j|$ we can consider the associated doubled SymTFT with Hilbert space $\mathcal{H}\otimes \overline{\mathcal{H}}$ following subsection \ref{ssec:WignerAsWave}. The resulting SymTFT is the direct product of the original SymTFT with Hilbert space $\mathcal{H}$ and its conjugate. A mixed state boundary condition
\begin{equation}
\label{eq:mixedelectric}
\widehat{\rho}_B=\sum_{ij}{b}_{ij}|e_i\rangle \langle e_j|\,,
\end{equation}
formulated here with respect to an electric polarization, then corresponds to a classical function $p^{[r,s]}_B(e,m)$ on phase space defined by using the discrete analog of \eqref{eq:greatexpectations},
\begin{equation}
\text{Tr}\lb \widehat{\rho}_B\widehat{\rho}_Z \rb=\sum_{e,m} p^{[r,s]}_B(e,m)W_Z^{[r,s]}(e,m)\,.
\end{equation}
This allows us to define the SymTFT / SymTh boundary condition
\begin{equation}
 \langle \!\langle \rho_B^{[r,s]}| \equiv \sum_{\mathbb{B}} p_B(\mathbb{B})  \langle \!\langle \mathbb{B}^{[r,s]}|
\end{equation}
with phase space coordinate $\mathbb{B}=(e,m)$. The key point is now that this boundary condition does not depend on any choice of polarization. In principle, $p_B(\mathbb{B}) $ may be any classical function on phase space\footnote{Observe that starting from a theory admitting a polarization and a density matrix $\widehat{\rho}_B$ as in \eqref{eq:mixedelectric}, we produce a function depending on $(\text{dim}\,\mathcal{H})-1$ variables. Generic functions on phase space depend on  $(\text{dim}\,\mathcal{H})^2$ parameters.} and we can use the above expressions to associate a boundary condition to it. Moreover, the above expression continues to hold with equal utility when $\mathbb{B}$ are drawn from a configuration space which does not split into position and momentum states, i.e., when no polarization exists.

Given the above boundary condition, we then have the overlap
\begin{equation}
\label{eq:WignerAsWave}
\langle \!\langle \rho_B^{[r,s]} | \rho_Z \rangle \!\rangle=\sum_{\mathbb{B}}p_B(\mathbb{B})W_Z^{[r,s]} (\mathbb{B})\,.
\end{equation}
In particular, Wigner's function
\begin{equation}
 \langle \!\langle \mathbb{B}^{[r,s]}|  \rho_Z \rangle \!\rangle=W_Z^{[r,s]} (\mathbb{B})
\end{equation}
is interpreted as specifying a natural set of partition functions to a mixed state $\widehat{\rho}_Z$ of relative theories. Further, $\langle \!\langle \rho_B^{[r,s]} | \rho_Z \rangle \!\rangle$ is then understood as a weighted sum of this basis of partition functions with weights specified by the classical distribution $p_B(\mathbb{B})$. Here, the index on $p_B$ only makes reference to the ``Boundary".

The above constructions rely on the doubled SymTFT with Hilbert space $\mathcal{H}\otimes \overline{\mathcal{H}}$. Interestingly, given the sandwich corresponding to $\langle \!\langle \rho_B^{[r,s]} | \rho_Z \rangle \!\rangle$, we can ``unfold it" and represent it as a sum over a disjoint family of sandwiches of the original SymTFT with Hilbert space $\mathcal{H}$. See also \cite{Ma:2024kma} for related discussion.

\begin{figure}
\centering
\scalebox{0.8}{
\begin{tikzpicture}
	\begin{pgfonlayer}{nodelayer}
		\node [style=BigCircleBlue] (0) at (-6.5, 0) {};
		\node [style=BigCircleRed] (1) at (-3.5, 0) {};
		\node [style=none] (2) at (-5, -1.75) {(i)};
		\node [style=none] (3) at (-3.5, 0.65) {$|\rho_Z\rangle\!\rangle$};
		\node [style=none] (4) at (-5.15, 0.15) {};
		\node [style=none] (5) at (-4.85, -0.15) {};
		\node [style=none] (6) at (-3.5, -0.15) {};
		\node [style=none] (7) at (-6.5, -0.15) {};
		\node [style=none] (8) at (-6.5, 0.15) {};
		\node [style=none] (9) at (-3.5, 0.15) {};
		\node [style=none] (10) at (-6.5, 0.70) {$\langle \!\langle \mathbb{B}^{[r,s]}|$};
		\node [style=BigCircleRed] (12) at (1.5, 0) {};
		\node [style=none] (13) at (0, -1.75) {(ii)};
		\node [style=none] (14) at (1.5, 0.65) {$|\rho_Z\rangle\!\rangle$};
		\node [style=none] (15) at (-0.15, 0.15) {};
		\node [style=none] (16) at (0.15, -0.15) {};
		\node [style=none] (17) at (1.5, -0.15) {};
		\node [style=none] (18) at (-1.5, -0.15) {};
		\node [style=none] (19) at (-1.5, 0.15) {};
		\node [style=none] (20) at (1.5, 0.15) {};
		\node [style=none] (21) at (-2.125, 0.05) {Id};
		\node [style=none] (24) at (11, -1.75) {(iv)};
		\node [style=none] (33) at (0, -2.5) {};
		\node [style=BigCircleRed] (34) at (9.5, 0) {};
		\node [style=BigCircleRed] (35) at (12.5, 0) {};
		\node [style=SmallCircleGreen] (36) at (11, 0) {};
		\node [style=SmallCircleGreen] (37) at (-0.75, -0.15) {};
		\node [style=SmallCircleGreen] (38) at (-0.75, 0.15) {};
		\node [style=BigCircleWhite] (39) at (-1.5, 0) {};
		\node [style=none] (40) at (9.5, 0.65) {$\langle Z_j|$};
		\node [style=none] (41) at (12.5, 0.65) {$| Z_i\rangle$};
		\node [style=BigCircleRed] (42) at (6.5, 0) {};
		\node [style=none] (43) at (5, -1.75) {(iii)};
		\node [style=none] (44) at (6.5, 0.65) {$|\rho_Z\rangle\!\rangle$};
		\node [style=none] (45) at (4.85, 0.15) {};
		\node [style=none] (46) at (5.15, -0.15) {};
		\node [style=none] (47) at (6.5, -0.15) {};
		\node [style=none] (48) at (3.5, -0.15) {};
		\node [style=none] (49) at (3.5, 0.15) {};
		\node [style=none] (50) at (6.5, 0.15) {};
		\node [style=SmallCircleGreen] (53) at (4.25, -0.15) {};
		\node [style=SmallCircleGreen] (54) at (4.25, 0.15) {};
		\node [style=none] (55) at (8.5, 0) {$\sum_{ij}z_{ij}$};
		\node [style=none] (56) at (-0.75, 0.75) {$\mathbb{U}_{B}^{[s]}$};
		\node [style=none] (57) at (-0.75, -0.75) {$\mathcal{C} \:\! \mathbb{U}_{B}^{[r] \, \dagger}$};
		\node [style=none] (58) at (11.125, -0.625) {$\mathbb{U}_{B}^{[s]}\:\!\mathcal{C} \:\! \mathbb{U}_{B}^{[r] \, \dagger}$};
	\end{pgfonlayer}
	\begin{pgfonlayer}{edgelayer}
		\draw [style=ThickLine] (5.center) to (6.center);
		\draw [style=ThickLine] (4.center) to (8.center);
		\draw [style=ArrowLineRight] (9.center) to (4.center);
		\draw [style=ArrowLineRight] (7.center) to (5.center);
		\draw [style=ThickLine] (16.center) to (17.center);
		\draw [style=ThickLine] (15.center) to (19.center);
		\draw [style=ArrowLineRight] (20.center) to (15.center);
		\draw [style=ArrowLineRight] (18.center) to (16.center);
		\draw [style=ThickLine] (34) to (35);
		\draw [style=ThickLine] (46.center) to (47.center);
		\draw [style=ThickLine] (45.center) to (49.center);
		\draw [style=ArrowLineRight] (50.center) to (45.center);
		\draw [style=ArrowLineRight] (48.center) to (46.center);
		\draw [style=ThickLine, bend right=270, looseness=2.00] (48.center) to (49.center);
	\end{pgfonlayer}
\end{tikzpicture}}
\caption{We sketch steps relating ``unfolding" the doubled SymTFT sandwich to a sum of SymTFT sandwiches with two physical boundary conditions. (i): Starting point, i.e., the initial doubled configuration with partition function given by Wigner's function. Boundary conditions are determined by two pure states. (ii): We separate out two operators from the boundary condition, which are indicated by green dots and are supported in one of the two systems tensored to give the overall doubled system. (iii): The boundary condition trivializes to the identity operator $\text{Id}$ (with respect to a single copy of the SymTFT) upon separating out these two operators. We can therefore reconnect the two SymTFT supports to give a single connected system. (iv): Finally, we make $\rho_Z=\sum_{ij}z_{ij}|Z_i\rangle \langle Z_j|$ explicit and fuse the two green dots to $\mathbb{U}_{B}^{[s]}\:\!\mathcal{C} \:\! \mathbb{U}_{B}^{[r] \, \dagger}$.}
\label{fig:OpeningUp}
\end{figure}
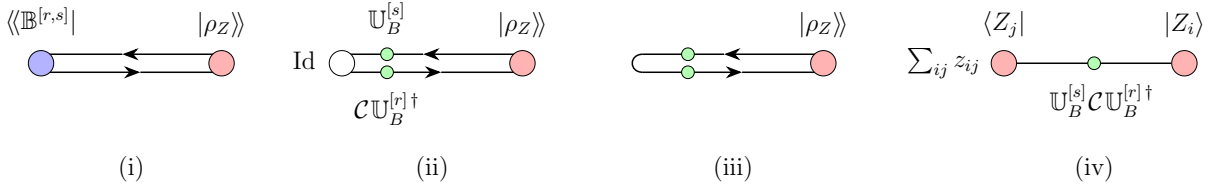

In order to demonstrate this, let us momentarily specialize to the pure state $\widehat{\rho}=\left\vert
Z\right\rangle \left\langle Z\right\vert $. Then, we have:
\begin{equation}\label{eq:NoPolarization}
W^{[r,s]}_{Z}(\mathbb{B})=\langle \! \langle \mathbb{B}^{[r,s]}|\rho_Z\rangle \!\rangle=\left\langle Z\right\vert \mathbb{U}^{[s]}_{\mathbb{B}}\:\!\mathcal{C}\:\!%
\mathbb{U}_{\mathbb{B}}^{[r] \, \dag}\left\vert Z\right\rangle .
\end{equation}
This quantity expresses Wigner's function as a matrix element of the original SymTFT. Further, both boundaries are physical, avoiding reference to a polarization. In the SymTFT / SymTh formalism, we can view the $\mathbb{U}_{B}$'s as $(-1)$-form symmetry operators which modify
the choice of boundary conditions (see figure \ref{eq:AvoidPolarization}). More generally, with $\widehat{\rho}_Z=\sum_{ij}z_{ij}|Z_i\rangle \langle Z_j|$ it follows that:
\begin{equation}
\label{eq:WignerAsWave2}
\langle \!\langle \rho_B^{[r,s]} | \rho_Z \rangle \!\rangle=\sum_{\mathbb{B},ij}z_{ij} p_B(\mathbb{B})\left\langle Z_j\right\vert \mathbb{U}^{[s]}_{\mathbb{B}}\:\!\mathcal{C}\:\!%
\mathbb{U}_{\mathbb{B}}^{[r] \, \dag}\left\vert Z_i\right\rangle \,.
\end{equation}
See figure \ref{fig:OpeningUp} for a depiction.

Let us contrast the above to situations with an absolute polarization and pure state $\widehat{\rho}_Z=|Z\rangle \langle Z|$ for which partition functions $Z(e)=\langle e|Z\rangle$ and $Z(m)= \langle m|Z\rangle$ are well-defined. Here, we
recover the expected partition function squared starting from \eqref{eq:NoPolarization}. For example, summing over the
magnetic degrees of freedom, we have:%
\begin{equation}
\underset{m}{\sum}W_{Z}^{[r,s]}(e,m)=\left\vert Z(e)\right\vert ^{2}.
\end{equation}
Similarly, we can still calculate expectation values of operators by
integrating over the entire classical phase space using
(\ref{eq:greatexpectations})
\begin{equation}
\left\langle Z\right\vert \widehat{G}\left\vert Z\right\rangle
=\underset{e,m}{\sum}W_{Z}^{[r,s]}(e,m)g(e,m).
\end{equation}
For example, the electric polarization projection operator $\widehat{E}_{e}=\left\vert
e\right\rangle \left\langle e\right\vert$ yields:%
\begin{equation}
\left\vert Z(e)\right\vert ^{2}=\underset{e',m'}{\sum}W_{Z}^{[r,s]}(e',m')E_{e}(e',m').
\end{equation}
It is evident that $E_{e}(e',m') = \delta_{e, e'}$.

Let us also make a comment regarding the application of the formalism of Wigner's function in the SymTFT / SymTh framework in expressions such as the trace:\footnote{That is, we trace over the space of states intrinsically defined by the SymTFT / SymTh treated as a QFT in its own right. Of course, one must enrich the space of state to include the physical states such as $\vert Z \rangle$.}
\begin{equation}
W_{Z}^{[r,s]}(\mathbb{B}) \sim \mathrm{Tr}_{\mathrm{gap/free}}(\widehat{\rho}_Z \mathbb{U}^{[s]}_{\mathbb{B}}\:\!\mathcal{C}\:\!%
\mathbb{U}_{\mathbb{B}}^{[r] \, \dag} )=\mathrm{Tr}_{\mathcal{H}}(\widehat{\rho}_Z \mathbb{U}^{[s]}_{\mathbb{B}}\:\!\mathcal{C}\:\!%
\mathbb{U}_{\mathbb{B}}^{[r] \, \dag}) ,
\end{equation}
with $\mathbb{B}$ a choice of field configuration in the phase space. Here the trace is actually restricted to those states associated with gapped / free boundary conditions, i.e., elements of $\mathcal{H}$, and of course does not cover for example enriched Neumann boundary conditions such as $|Z\rangle$. One can dispense with this restriction by instead inserting a suitable projection operator.

Finally,\footnote{We thank the referee for an insightful question leading to the addition of this paragraph.} note that Wigner's function allows us to quantify how complex, from a quantum information theoretic perspective, the mixed state $| \rho_Z \rangle \!\rangle$ is. For this, return to the continuum Wigner function in \eqref{eq:Wquasifunc} and recall that there, by a theorem of Hudson's \cite{HUDSON1974249}, one has that $W_\rho(p,q)\geq 0$ for a pure state $\rho=|\psi\rangle \langle \psi|$ only if $\psi(x)$ is Gaussian. The discrete analog of this result when Hilbert spaces are finite dimensional has $|\psi\rangle$ be a so-called stabilizer state \cite{Gross:2006wkl}. Generalizing to mixed states, Wigner functions are non-negative whenever $\rho= \sum_{i} p_i |\psi_i\rangle \langle \psi_i |$ with $|\psi_i\rangle$ stabilizer states. Another interesting result relating positivity to complexity is that Gaussian convolution of arbitrary quantum states eventually produces a strictly positive Wigner function, i.e., coarse-graining makes Wigner functions positive \cite{PhysRevA.44.R2775}. In contrast, magic states, in the sense of \cite{Kitaev:1997wr,PhysRevA.71.022316}, are pure states that are not stabilizer states and cannot be obtained by topological operations from stabilizer states. While exploring the characterization and consequences of negative Wigner functions for our setup of mixed states of partition vectors is beyond the scope of this work, we comment here that we expect these to organize orbits with respect to $(-1)$-form symmetry group actions of the underlying quantum system (as topological codimension-1 interfaces of a SymTFT may be interpreted as a $(-1)$-form symmetry operators with respect to its edge mode system) and leave further investigation for future work.

\subsection{Defects and Symmetry Operators}

\begin{figure}[t]
\centering
\scalebox{0.7}{
\begin{tikzpicture}
	\begin{pgfonlayer}{nodelayer}
		\node [style=none] (0) at (-5, -2) {};
		\node [style=none] (1) at (-5, 0.5) {};
		\node [style=none] (2) at (-4, 2) {};
		\node [style=none] (3) at (-4, -0.5) {};
		\node [style=none] (4) at (4, -2) {};
		\node [style=none] (5) at (4, 0.5) {};
		\node [style=none] (6) at (5, 2) {};
		\node [style=none] (7) at (5, -0.5) {};
		\node [style=none] (8) at (-0.5, -2) {};
		\node [style=none] (9) at (-0.5, 0.5) {};
		\node [style=none] (10) at (0.5, -0.5) {};
		\node [style=none] (11) at (0.5, 2) {};
		\node [style=none] (23) at (0, 2.75) {$\mathbb{U}_{\mathbb{B}}^{[s]}\:\! \mathcal{C} \:\!\mathbb{U}_{\mathbb{B}}^{[r],\dagger}$};
		\node [style=none] (24) at (-5, -7) {};
		\node [style=none] (25) at (-5, -4.5) {};
		\node [style=none] (26) at (-4, -3) {};
		\node [style=none] (27) at (-4, -5.5) {};
		\node [style=none] (28) at (4, -7) {};
		\node [style=none] (29) at (4, -4.5) {};
		\node [style=none] (30) at (5, -3) {};
		\node [style=none] (31) at (5, -5.5) {};
		\node [style=none] (32) at (-0.5, -7) {};
		\node [style=none] (33) at (-0.5, -4.5) {};
		\node [style=none] (34) at (0.5, -5.5) {};
		\node [style=none] (35) at (0.5, -3) {};
		\node [style=none] (37) at (-5, -12) {};
		\node [style=none] (38) at (-5, -9.5) {};
		\node [style=none] (39) at (-4, -8) {};
		\node [style=none] (40) at (-4, -10.5) {};
		\node [style=none] (41) at (4, -12) {};
		\node [style=none] (42) at (4, -9.5) {};
		\node [style=none] (43) at (5, -8) {};
		\node [style=none] (44) at (5, -10.5) {};
		\node [style=none] (45) at (-0.5, -12) {};
		\node [style=none] (46) at (-0.5, -9.5) {};
		\node [style=none] (47) at (0.5, -10.5) {};
		\node [style=none] (48) at (0.5, -8) {};
		\node [style=none] (50) at (-4.5, 0) {};
		\node [style=none] (51) at (4.5, 0) {};
		\node [style=none] (52) at (0, 0) {};
		\node [style=Circle] (53) at (0, -0.1) {};
		\node [style=Circle] (54) at (4.5, 0) {};
		\node [style=Circle] (55) at (-4.5, 0) {};
		\node [style=none] (56) at (2.5, -5) {};
		\node [style=none] (57) at (3, -5) {};
		\node [style=none] (58) at (2.75, -4.625) {};
		\node [style=none] (59) at (2.75, -5.375) {};
		\node [style=none] (60) at (2, -5) {};
		\node [style=none] (61) at (1, -5) {};
		\node [style=none] (62) at (-2.75, -10) {};
		\node [style=none] (63) at (-2.25, -10) {};
		\node [style=none] (64) at (-2.5, -9.625) {};
		\node [style=none] (65) at (-2.5, -10.375) {};
		\node [style=none] (66) at (0, -9.625) {};
		\node [style=none] (67) at (0, -10.375) {};
		\node [style=none] (68) at (0.25, -10) {};
		\node [style=none] (69) at (-0.25, -10) {};
		\node [style=none] (70) at (-7.5, 0) {(i)};
		\node [style=none] (71) at (-7.5, -5) {(ii)};
		\node [style=none] (72) at (-7.5, -10) {(iii)};
		\node [style=none] (73) at (0, -12.5) {};
		\node [style=SmallCircleBrown] (74) at (2.5, -5) {};
		\node [style=SmallCircleBrown] (75) at (-2.5, -10) {};
		\node [style=Circle] (76) at (0, -10) {};
		\node [style=none] (77) at (-5.75, 0) {$\langle Z| $};
		\node [style=none] (78) at (-5.75, -5) {$\langle Z| $};
		\node [style=none] (79) at (-5.75, -10) {$\langle Z| $};
		\node [style=none] (77) at (5.75, 0) {$| Z\rangle $};
		\node [style=none] (78) at (5.75, -5) {$| Z\rangle $};
		\node [style=none] (79) at (5.75, -10) {$ |Z\rangle $};
	\end{pgfonlayer}
	\begin{pgfonlayer}{edgelayer}
		\filldraw[fill=green!30, draw=green!30]  (-.5, -2) -- (-0.5, 0.5) -- (0.5, 2) -- (0.5, -0.5) -- cycle;
		\filldraw[fill=red!30, draw=red!30]  (4, -2) -- (4, 0.5) -- (5, 2) -- (5, -0.5) -- cycle;
		\filldraw[fill=red!30, draw=red!30]  (-5, -2) -- (-5, 0.5) -- (-4, 2) -- (-4, -0.5) -- cycle;
		\filldraw[fill=green!30, draw=green!30]  (-.5, -7) -- (-0.5, -4.5) -- (0.5, -3) -- (0.5, -5.5) -- cycle;
		\filldraw[fill=red!30, draw=red!30]  (4, -7) -- (4, -4.5) -- (5, -3) -- (5, -5.5) -- cycle;
		\filldraw[fill=red!30, draw=red!30]  (-5, -7) -- (-5, -4.5) -- (-4, -3) -- (-4, -5.5) -- cycle;
		\filldraw[fill=green!30, draw=green!30]  (-.5, -12) -- (-0.5, -9.5) -- (0.5, -8) -- (0.5, -10.5) -- cycle;
		\filldraw[fill=red!30, draw=red!30]  (4, -12) -- (4, -9.5) -- (5, -8) -- (5, -10.5) -- cycle;
		\filldraw[fill=red!30, draw=red!30]  (-5, -12) -- (-5, -9.5) -- (-4, -8) -- (-4, -10.5) -- cycle;
		\draw [style=ThickLine] (0.center) to (4.center);
		\draw [style=ThickLine] (5.center) to (1.center);
		\draw [style=ThickLine] (2.center) to (6.center);
		\draw [style=DottedLine] (3.center) to (7.center);
		\draw (9.center) to (11.center);
		\draw (11.center) to (10.center);
		\draw (10.center) to (8.center);
		\draw (8.center) to (9.center);
		\draw (1.center) to (0.center);
		\draw (1.center) to (2.center);
		\draw (4.center) to (7.center);
		\draw (7.center) to (6.center);
		\draw (6.center) to (5.center);
		\draw (5.center) to (4.center);
		\draw [style=DottedLine] (0.center) to (3.center);
		\draw [style=DottedLine] (3.center) to (2.center);
		\draw [style=ThickLine] (24.center) to (28.center);
		\draw [style=ThickLine] (29.center) to (25.center);
		\draw [style=ThickLine] (26.center) to (30.center);
		\draw [style=DottedLine] (27.center) to (31.center);
		\draw (33.center) to (35.center);
		\draw (35.center) to (34.center);
		\draw (34.center) to (32.center);
		\draw (32.center) to (33.center);
		\draw (25.center) to (24.center);
		\draw (25.center) to (26.center);
		\draw (28.center) to (31.center);
		\draw (31.center) to (30.center);
		\draw (30.center) to (29.center);
		\draw (29.center) to (28.center);
		\draw [style=DottedLine] (24.center) to (27.center);
		\draw [style=DottedLine] (27.center) to (26.center);
		\draw [style=ThickLine] (37.center) to (41.center);
		\draw [style=ThickLine] (42.center) to (38.center);
		\draw [style=ThickLine] (39.center) to (43.center);
		\draw [style=DottedLine] (40.center) to (44.center);
		\draw (46.center) to (48.center);
		\draw (48.center) to (47.center);
		\draw (47.center) to (45.center);
		\draw (45.center) to (46.center);
		\draw (38.center) to (37.center);
		\draw (38.center) to (39.center);
		\draw (41.center) to (44.center);
		\draw (44.center) to (43.center);
		\draw (43.center) to (42.center);
		\draw (42.center) to (41.center);
		\draw [style=DottedLine] (37.center) to (40.center);
		\draw [style=DottedLine] (40.center) to (39.center);
		\draw [style=PurpleLine, snake it] (50.center) to (51.center);
		%\draw [style=BrownLine, in=-180, out=90] (56.center) to (58.center);
		%\draw [style=BrownLine, in=90, out=0] (58.center) to (57.center);
		%\draw [style=BrownLine, in=0, out=-90] (57.center) to (59.center);
		%\draw [style=BrownLine, in=-90, out=180] (59.center) to (56.center);
		\draw [style=ArrowLineRight] (60.center) to (61.center);
		%\draw [style=BrownLine, in=-180, out=90] (62.center) to (64.center);
		%\draw [style=BrownLine, in=90, out=0] (64.center) to (63.center);
		%\draw [style=BrownLine, in=0, out=-90] (63.center) to (65.center);
		%\draw [style=BrownLine, in=-90, out=180] (65.center) to (62.center);
		%\draw [style=GreenLine, in=-180, out=90] (69.center) to (66.center);
		%\draw [style=GreenLine, in=90, out=0] (66.center) to (68.center);
		%\draw [style=GreenLine, in=0, out=-90] (68.center) to (67.center);
		%\draw [style=GreenLine, in=-90, out=180] (67.center) to (69.center);
		%\draw [style=GreenLine] (64.center) to (66.center);
		%\draw [style=GreenLine] (67.center) to (65.center);
		\draw [style=DashedLine] (75.center) to (76.center);
	\end{pgfonlayer}
\end{tikzpicture}
}
\caption{We sketch defect and symmetry operators in presence of a $\mathbb{U}_{\mathbb{B}}^{[s]}\:\! \mathcal{C} \:\!\mathbb{U}_{\mathbb{B}}^{[r],\dagger}$ insertion viewed as a codimension-one SymTFT operator / wall. (i): Defect operators piercing this wall may be dressed along their intersection with it (black dot). (ii) and (iii): Deforming a symmetry operator across the wall, it may remain attached thereto (dashed line). When genuine operators are mapped onto genuine operators, the dashed line is absent / trivial.  }
\label{fig:UPhiInteraction}
\end{figure}
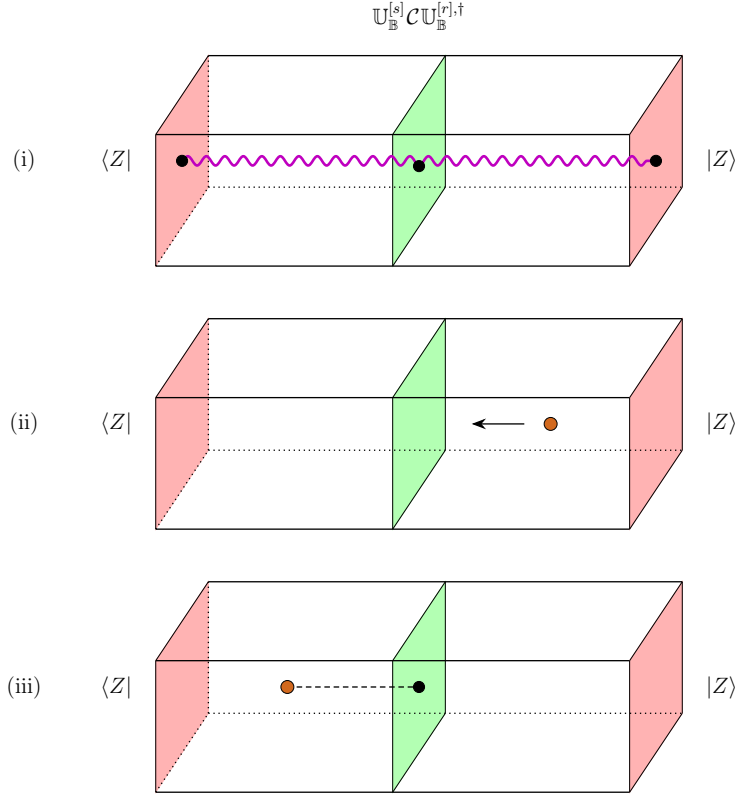

One of the important features of the SymTFT framework is that it provides a systematic way to organize the symmetry data of a system. In particular, one has the topological symmetry operators that link with defects of the theory. Recall that a defect stretches between the physical and gapped / free boundaries, whereas a symmetry operator is one that topologically links with this defect. Now, the precise spectrum of genuine
defects and symmetry operators are those that can be specified without regard to any higher-dimensional operators. By contrast, non-genuine
defects and symmetry operators are those that are better viewed as the boundaries of a higher-dimensional defect / operator.

How is this information taken into account by Wigner's function? To begin, consider the case where the boundary and QFT state are both pure, i.e., $\widehat{\rho}_B = \vert B \rangle \langle B \vert$ and $\widehat{\rho}_Z = \vert Z \rangle \langle Z \vert$. We consider the effects of the operator $\mathbb{U}_{\mathbb{B}}^{[t]}$ used to specify the background fields in the first place. This can be viewed as specifying a codimension-one defect in the bulk, i.e., a wall. From the perspective of the $D$-dimensional QFT, it fills all of the spacetime. Suppose we now consider a defect which stretches along the auxiliary bulk direction (see subfigure (i) of figure \ref{fig:UPhiInteraction}). As it passes through the $\mathbb{U}_{\mathbb{B}}^{[t]}$ wall, it can be dressed by additional operators / defects. This dressing is the analog of making a defect genuine versus non-genuine. By the same token, a symmetry operator which crosses through the $\mathbb{U}_{\mathbb{B}}^{[t]}$ wall may either pass through, remaining detached from the wall but possibly changing to a different operator, or may instead attach back to the wall via a ``flux tube" (see subfigures (ii) and (iii) of figure \ref{fig:UPhiInteraction}). The fully detached case corresponds to a genuine symmetry operator, and the case that attaches back to the wall is a non-genuine symmetry operator.

Now that we have explained how this works in the case of pure states, we can extend these considerations to more general mixed states of relative QFTs $\widehat{\rho}_Z=\sum_i z_i |Z_i\rangle \langle Z_i|$ and mixed boundary conditions, as specified by the classical function $p_{B}(\mathbb{B})$. Returning to (\ref{eq:WignerAsWave2}), recall that we have the overlap:
\begin{equation}
\langle \!\langle \rho_B^{[r,s]} | \rho_Z \rangle \!\rangle=\sum_{\mathbb{B},i}z_{i} p_B(\mathbb{B})\left\langle Z_i\right\vert \mathbb{U}^{[s]}_{\mathbb{B}}\:\!\mathcal{C}\:\!%
\mathbb{U}_{\mathbb{B}}^{[r] \, \dag}\left\vert Z_i\right\rangle \,.
\end{equation}
Given a linking configuration of a defect $\mathcal{D}$ and symmetry operator $\mathcal{U}$, we can compute
$\langle \!\langle \rho_B^{[r,s]} \vert \mathrm{lnk}(\mathcal{D}, \mathcal{U}) \vert \rho_Z \rangle \!\rangle$ as follows.
Inserting defects and symmetry operators follows by working with each summand individually, i.e., we evaluate expectation values
with defects and symmetry operators (genuine or otherwise) inserted. We then perform a weighted sum of these answers with weights $p_{B}(\mathbb{B}) z_i$.

\subsection{Illustrative Example: 6D SCFTs}

To illustrate some of these points, consider a 6D superconformal field theory (SCFT) with non-trivial defect group \cite{DelZotto:2015isa} with $\mathcal{N}=(2,0)$ supersymmetry and labelled by some ADE Lie algebra.\footnote{See \cite{Heckman:2018jxk, Argyres:2022mnu} for recent  reviews of 6D SCFTs.} Similar considerations hold for $\mathcal{N} = (1,0)$ SCFTs, but in this case, one must also typically contend with possible 1-form and 0-form symmetries (see e.g., \cite{Hubner:2022kxr, Cvetic:2022imb, Heckman:2022suy}). For ease of exposition, we leave this extension implicit in what follows.

In the SymTFT formalism, treating the 2-form symmetries is rather awkward because there is often no polarization available for the defect group unless one makes a number of restrictive choices.\footnote{For example, if the defect group splits as $\mathbb{D} = L \oplus \overline{L}$ with $L$ a Lagrangian subgroup, then an absolute polarization can be produced, but there is no guarantee that this is always possible. For further discussion on this point, see reference \cite{Lawrie:2023tdz}.} Rather, one typically has to contend with a vector of partition functions. Of course, the dot product of two vectors returns a scalar. As such, we expect Wigner's function to make sense even in these situations.

In this case, the bulk SymTFT for the 2-form symmetries is governed by a Chern-Simons-like action for abelian 3-form $C^I$ potentials
of the form (see e.g., \cite{Heckman:2017uxe, Apruzzi:2022dlm, Baume:2023kkf, GarciaEtxebarria:2024fuk}):
\begin{equation}
\mathcal{S}_{7D} = \frac{K_{IJ}}{4 \pi} \int_{7D} C^{I} d C^{J}\,,
\end{equation}
with integer level matrix $K_{IJ}\in \Z$ which is minus the Cartan matrix
of the ADE Lie algebra associated to the 6D SCFT.\footnote{I.e., the intersection form of two-cycles for the resolution of the ADE singularity
$\mathbb{C}^{2} / \Gamma_{ADE}$ for $\Gamma_{ADE} \subset SU(2)$ a finite subgroup.}
The defect group \cite{DelZotto:2015isa,} is then the finite group
\begin{equation}\label{eq:DefectGroup6D}
\mathbb{D}\cong \text{Coker}\,K_{IJ}\,,
\end{equation}
and equipped with a Dirac pairing induced by the inverse $K^{IJ}$. These SCFTs, on their tensor branch, famously contain 2-form gauge fields with anti-selfdual curvatures, and consequently, there is no distinction between electric and magnetic defects, all defects are subsumed into $\mathbb{D}$. In structure is extremely similar to the Chern-Simons example discussed in section \ref{sec:ABCS} and therefore computations will be essentially analogous, with only core difference being that we will focus on non-compact flat spacetimes.

When the order $|\mathbb{D}|$ is not a square, then we cannot specify a polarization. For example, all SCFTs associated labelled by the Lie algebra $\mathfrak{su}_N$ with $N$ not a perfect square are intrinsically relative in this sense. Nonetheless, the SymTFT with action $ \mathcal{S}_{7D} $ makes sense, and the SCFT realizes a physical boundary condition to it. See, for example, \cite{Gukov:2020btk} for further discussion.

Further, bulk topological operators can then be built from general periods of the $C^{I}$:
\begin{equation}
\mathcal{T}_{\nu, \Sigma} = \exp \left( i \int_{\Sigma} \nu_I \widehat{C}^{I} \right),
\end{equation}
where $\Sigma$ is a 3-manifold and the hatted variable indicates the operator associated to field $C$ by canonical quantization, similar to our discussion of Abelian Chern-Simons theory in section \ref{sec:ABCS}. These satisfy the braiding relations:\footnote{See also the discussion in \cite{DelZotto:2015isa, Apruzzi:2016nfr, Heckman:2017uxe, Gukov:2020btk, Apruzzi:2022dlm}.}
\begin{equation}
\mathcal{T}_{\nu, \Sigma} \mathcal{T}_{\nu^{\prime}, \Sigma^{\prime}} = \exp\left(2 \pi i \times \nu_{I} \left(\frac{1}{K}\right)^{IJ} \nu^{\prime}_{J} \times \mathrm{lnk}(\Sigma, \Sigma^{\prime}) \right)\mathcal{T}_{\nu^{\prime}, \Sigma^{\prime}} \mathcal{T}_{\nu, \Sigma}.
\end{equation}
Here, $\mathrm{lnk}(\Sigma_1, \Sigma_2)$ is the Gaussian linking between the 3-surfaces in the 7D SymTFT worldvolume.
Observe that we can also split these operators up into defects and symmetry operators of the 6D SCFT, depending on whether they extend in the auxiliary bulk direction (as expected for the defects) or do not (as expected for the symmetry operators) of the SymTFT.

Nevertheless, it is clear that we can still specify a Wigner's function for the 2-form symmetries. To start, we comment that although the phase space of the 7D SymTFT depends on periods of the 3-form potential $C$, screening effects in the 6D SCFT tell us that in flat spacetimes it suffices to label the
phase space operators as $\mathbb{U}_{\mu}^{[s]}$ with $\mu \in \mathbb{D}$ an element in the defect group \cite{DelZotto:2015isa}. We will set $s=1$ going forward. 

Indeed, given a $\mu \in \mathbb{D}$, we can produce a representative three-form potential $\mu_I=K_{IJ}C^{J}$ which is dual to the non-compact 3-cycle $\mathbb{R}^3$ in the flat 6D spacetime. Any two choices of such coordinate planes are gauge equivalent. The gauge invariant data of this background field is packaged in terms of its periods, as obtained by integrating over a basis of three-cycles. Owing to screening by dynamical states (see \cite{DelZotto:2015isa}), these periods are captured by elements $\mu \in \mathbb{D}$. By abuse of notation, we can then construct the operator:
\begin{equation}
\mathbb{U}_{\mu} = \exp\left(  -iK_{IJ}\int_{6D}
C^I  \widehat{C}^{J}\right)  ,
\end{equation}
with 3-form chemical potential $C$, namely the phase is defined by integrating over the entire 6D spacetime.  This expression should be compared to \eqref{eq:Important}.

Next, consider a pure state $\widehat{\rho}_Z=|Z\rangle \langle Z|$. Then we have Wigner's function
\begin{equation}
W_Z(\mu)=\langle \!\langle \mu | \rho_Z\rangle \!\rangle=\langle Z|  \mathbb{U}_{\mu}\:\! \mathcal{C}\:\! \mathbb{U}_{\mu}^{\dag} |Z\rangle\,,
\end{equation}
which is the partition function of a doubled system of 6D SCFTs (which is absolute) with 2-form symmetry group
\begin{equation}
\Gamma^{(2\text{-form})}\cong \mathbb{D}^\vee\,,
\end{equation}
where $\vee$ indicates the Pontryagin dual group. Since $\mathbb{D}$ is an abelian group, we have
the non-canonical isomorphism $ \mathbb{D}^\vee\cong \mathbb{D}$.

\section{Compactification and Holography}
\label{sec:4}

These considerations become especially helpful in the context of systems where there is already a natural notion of a
$(D+1)$-dimensional bulk. In this section, we discuss two examples motivated by higher-dimensional gravity, namely string compactification, as well as holography. Of course, there are also examples that blend these considerations.

\subsection{String Compactification}

\begin{figure}[t]
\centering
\scalebox{0.7}{
\begin{tikzpicture}
	\begin{pgfonlayer}{nodelayer}
		\node [style=none] (0) at (7.5, 1.25) {};
		\node [style=none] (1) at (5.25, 2) {};
		\node [style=none] (2) at (5.25, 0.5) {};
		\node [style=Star] (3) at (7.5, 1.25) {};
		\node [style=none] (4) at (3.5, 0) {};
		\node [style=none] (5) at (3.5, -0.75) {};
		\node [style=none] (6) at (5.25, -1.25) {};
		\node [style=none] (7) at (1.75, -1.25) {};
		\node [style=BigCircleBlue] (8) at (1.75, -1.25) {};
		\node [style=BigCircleRed] (9) at (5.25, -1.25) {};
		\node [style=none] (10) at (5.25, -2) {$|Z\rangle$};
		\node [style=none] (11) at (1.75, -2) {$\langle B|$};
		\node [style=none] (12) at (3.5, -5) {\large (i)};
		\node [style=none] (13) at (13, 7.95) {};
		\node [style=none] (14) at (14, 7.95) {};
		\node [style=none] (15) at (14.25, 8) {};
		\node [style=none] (16) at (12.75, 8) {};
		\node [style=none] (17) at (15.75, 8) {};
		\node [style=none] (18) at (16, 7.95) {};
		\node [style=none] (19) at (17, 7.95) {};
		\node [style=none] (20) at (17.25, 8) {};
		\node [style=none] (21) at (15, 9) {};
		\node [style=none] (22) at (15, 7) {};
		\node [style=none] (23) at (16.5, 9.5) {};
		\node [style=none] (24) at (16.5, 6.5) {};
		\node [style=none] (25) at (13.5, 6.5) {};
		\node [style=none] (26) at (13.5, 9.5) {};
		\node [style=none] (27) at (11.5, 8) {};
		\node [style=none] (28) at (18.25, 8) {};
		\node [style=none] (29) at (12.25, 7.25) {};
		\node [style=none] (30) at (13, 7) {};
		\node [style=none] (31) at (12.25, 8.75) {};
		\node [style=none] (32) at (13, 9) {};
		\node [style=none] (33) at (17.5, 7.25) {};
		\node [style=none] (34) at (16.75, 7) {};
		\node [style=none] (35) at (16.75, 9) {};
		\node [style=none] (36) at (17.5, 8.75) {};
		\node [style=none] (37) at (17.5, 10) {};
		\node [style=none] (38) at (17.5, 6) {};
		\node [style=none] (39) at (12.25, 10) {};
		\node [style=none] (40) at (12.25, 6) {};
		\node [style=Star] (41) at (12.25, 10) {};
		\node [style=Star] (42) at (17.5, 10) {};
		\node [style=Star] (43) at (17.5, 6) {};
		\node [style=Star] (44) at (12.25, 6) {};
		%\node [style=none] (45) at (15, 4.75) {$\sim$};
		\node [style=none] (46) at (13.25, 3.5) {};
		\node [style=none] (47) at (13.25, 2.75) {};
		\node [style=none] (48) at (13.25, 2.25) {};
		\node [style=none] (49) at (13.25, 1.5) {};
		\node [style=none] (50) at (13.25, 1) {};
		\node [style=none] (51) at (13.25, 0.25) {};
		\node [style=none] (52) at (13.25, -0.25) {};
		\node [style=none] (53) at (13.25, -1) {};
		\node [style=none] (54) at (11.25, 1.75) {};
		\node [style=none] (55) at (11.5, 1.71) {};
		\node [style=none] (56) at (12.25, 1.71) {};
		\node [style=none] (57) at (12.5, 1.75) {};
		\node [style=none] (58) at (11.25, 0.75) {};
		\node [style=none] (59) at (11.5, 0.71) {};
		\node [style=none] (60) at (12.25, 0.71) {};
		\node [style=none] (61) at (12.5, 0.75) {};
		\node [style=none] (62) at (14, 3.5) {};
		\node [style=none] (63) at (14, 2.75) {};
		\node [style=none] (64) at (16, 3.5) {};
		\node [style=none] (65) at (16, 2.75) {};
		\node [style=none] (66) at (14, 2.25) {};
		\node [style=none] (67) at (14, 1.5) {};
		\node [style=none] (68) at (16, 2.25) {};
		\node [style=none] (69) at (16, 1.5) {};
		\node [style=none] (70) at (14, 1) {};
		\node [style=none] (71) at (14, 0.25) {};
		\node [style=none] (72) at (16, 1) {};
		\node [style=none] (73) at (16, 0.25) {};
		\node [style=none] (74) at (14, -0.25) {};
		\node [style=none] (75) at (14, -1) {};
		\node [style=none] (76) at (16, -0.25) {};
		\node [style=none] (77) at (16, -1) {};
		\node [style=none] (78) at (16.75, 3.5) {};
		\node [style=none] (79) at (16.75, 2.75) {};
		\node [style=none] (80) at (16.75, 2.25) {};
		\node [style=none] (81) at (16.75, 1.5) {};
		\node [style=none] (82) at (16.75, 1) {};
		\node [style=none] (83) at (16.75, 0.25) {};
		\node [style=none] (84) at (16.75, -0.25) {};
		\node [style=none] (85) at (16.75, -1) {};
		\node [style=none] (86) at (18.5, 3.125) {};
		\node [style=none] (87) at (18.5, 1.875) {};
		\node [style=none] (88) at (18.5, 0.625) {};
		\node [style=none] (89) at (18.5, -0.625) {};
		\node [style=Star] (90) at (18.5, 3.125) {};
		\node [style=Star] (91) at (18.5, 1.875) {};
		\node [style=Star] (92) at (18.5, 0.625) {};
		\node [style=Star] (93) at (18.5, -0.625) {};
		\node [style=none] (94) at (15, -1.5) {};
		\node [style=none] (95) at (15, -2.25) {};
		\node [style=none] (96) at (16.75, -2.75) {};
		\node [style=none] (97) at (13.25, -2.75) {};
		\node [style=BigCircleBlue] (98) at (13.25, -2.75) {};
		\node [style=BigCircleRed] (99) at (16.75, -2.75) {};
		\node [style=none] (100) at (16.75, -3.5) {$\otimes_i|Z_i\rangle$};
		\node [style=none] (101) at (13.25, -3.5) {$\langle B|$};
		\node [style=none] (102) at (15, -5) {\large (ii)};
		\node [style=none] (103) at (2.5, 2) {};
		\node [style=none] (104) at (2.5, 0.5) {};
		\node [style=none] (105) at (4.5, 2) {};
		\node [style=none] (106) at (4.5, 0.5) {};
		\node [style=none] (107) at (1.75, 2) {};
		\node [style=none] (108) at (1.75, 0.5) {};
		\node [style=none] (109) at (15, 6.25) {$X$};
		\node [style=none] (110) at (11.5, -1.25) {$X^{\text{bulk}}$};
		\node [style=none] (111) at (18, -1.5) {$X^{\text{loc}}$};
		\node [style=none] (112) at (15, 4) {$I\times K$};
		\node [style=none] (113) at (15, 5.5) {};
		\node [style=none] (114) at (15, 4.75) {};
		\node [style=none] (115) at (12.5, 9.25) {};
		\node [style=none] (116) at (17.25, 9.25) {};
		\node [style=none] (117) at (17.25, 6.75) {};
		\node [style=none] (118) at (12.5, 6.75) {};
		\node [style=none] (119) at (13.25, 3.25) {};
		\node [style=none] (120) at (13.25, 2) {};
		\node [style=none] (121) at (13.25, 0.75) {};
		\node [style=none] (122) at (13.25, -0.5) {};
		\node [style=none] (123) at (16.75, 3.25) {};
		\node [style=none] (124) at (16.75, 2) {};
		\node [style=none] (125) at (16.75, 0.75) {};
		\node [style=none] (126) at (16.75, -0.5) {};
	\end{pgfonlayer}
	\begin{pgfonlayer}{edgelayer}
		\filldraw[fill=blue!30, draw=none] (1.75,1.25) ellipse (0.23 and 0.75);
		\filldraw[fill=red!30, draw=none] (5.25, 1.25) ellipse (0.23 and 0.75);
		\filldraw[fill=blue!30, draw=none] (13.25, 3.125) ellipse (0.23 and 0.375);
		\filldraw[fill=blue!30, draw=none] (13.25, 1.875) ellipse (0.23 and 0.375);
		\filldraw[fill=blue!30, draw=none] (13.25, 0.625) ellipse (0.23 and 0.375);
		\filldraw[fill=blue!30, draw=none] (13.25, -0.625) ellipse (0.23 and 0.375);
		\filldraw[fill=red!30, draw=none] (16.75, 3.125) ellipse (0.23 and 0.375);
		\filldraw[fill=red!30, draw=none] (16.75, 1.875) ellipse (0.23 and 0.375);
		\filldraw[fill=red!30, draw=none] (16.75, 0.625) ellipse (0.23 and 0.375);
		\filldraw[fill=red!30, draw=none] (16.75, -0.625) ellipse (0.23 and 0.375);
		\filldraw[fill=red!30, draw=red!30] (16.75, 3.5) -- (16.75, 2.75) -- (18.5, 3.125) -- cycle;
		\filldraw[fill=red!30, draw=red!30] (16.75, 2.25) -- (16.75, 1.5) -- (18.5, 1.875) -- cycle;
		\filldraw[fill=red!30, draw=red!30] (16.75, 1) -- (16.75, 0.25) -- (18.5, 0.625) -- cycle;
		\filldraw[fill=red!30, draw=red!30] (16.75, -0.25) -- (16.75, -1) -- (18.5, -0.625) -- cycle;
		\filldraw[fill=red!30, draw=red!30] (5.25, 2) -- (5.25, 0.5) -- (7.5, 1.25) -- cycle;
		\draw [style=ThickLine] (2.center) to (0.center);
		\draw [style=ThickLine] (0.center) to (1.center);
		\draw [style=ThickLine, bend right=90, looseness=0.50] (1.center) to (2.center);
		\draw [style=ArrowLineRight] (4.center) to (5.center);
		\draw [style=ThickLine] (7.center) to (6.center);
		\draw [style=ThickLine, bend left=15] (13.center) to (14.center);
		\draw [style=ThickLine, bend left=15] (15.center) to (16.center);
		\draw [style=ThickLine, bend left=15] (18.center) to (19.center);
		\draw [style=ThickLine, bend left=15] (20.center) to (17.center);
		\draw [style=ThickLine, in=-180, out=0] (26.center) to (21.center);
		\draw [style=ThickLine, in=-180, out=0] (21.center) to (23.center);
		\draw [style=ThickLine, in=90, out=0] (23.center) to (28.center);
		\draw [style=ThickLine, in=0, out=-90] (28.center) to (24.center);
		\draw [style=ThickLine, in=0, out=180] (24.center) to (22.center);
		\draw [style=ThickLine, in=0, out=-180] (22.center) to (25.center);
		\draw [style=ThickLine, in=-90, out=-180] (25.center) to (27.center);
		\draw [style=ThickLine, in=-180, out=90] (27.center) to (26.center);
		\draw [style=ThickLine, in=45, out=-90, looseness=0.50] (39.center) to (31.center);
		\draw [style=ThickLine, in=-180, out=-60, looseness=0.50] (39.center) to (32.center);
		\draw [style=ThickLine, in=15, out=-120, looseness=0.50] (37.center) to (35.center);
		\draw [style=ThickLine, in=120, out=-90, looseness=0.50] (37.center) to (36.center);
		\draw [style=ThickLine, in=90, out=-135, looseness=0.50] (33.center) to (38.center);
		\draw [style=ThickLine, in=-15, out=120, looseness=0.50] (38.center) to (34.center);
		\draw [style=ThickLine, in=-165, out=60, looseness=0.50] (40.center) to (30.center);
		\draw [style=ThickLine, in=90, out=-45, looseness=0.50] (29.center) to (40.center);
		\draw [style=ThickLine, bend right=90, looseness=1.75] (47.center) to (48.center);
		\draw [style=ThickLine, in=180, out=-180, looseness=1.75] (49.center) to (50.center);
		\draw [style=ThickLine, bend right=90, looseness=1.75] (51.center) to (52.center);
		\draw [style=ThickLine, bend right=90, looseness=2.00] (46.center) to (53.center);
		\draw [style=ThickLine, bend left=15] (55.center) to (56.center);
		\draw [style=ThickLine, bend left=15] (57.center) to (54.center);
		\draw [style=ThickLine, bend left=15] (59.center) to (60.center);
		\draw [style=ThickLine, bend left=15] (61.center) to (58.center);
		\draw [style=ThickLine, bend right=90] (46.center) to (47.center);
		\draw [style=ThickLine, bend left=90] (49.center) to (48.center);
		\draw [style=ThickLine, bend right=90] (50.center) to (51.center);
		\draw [style=ThickLine, bend right=90] (52.center) to (53.center);
		\draw [style=ThickLine, bend left=90] (46.center) to (47.center);
		\draw [style=ThickLine, bend left=90] (48.center) to (49.center);
		\draw [style=ThickLine, bend left=90] (50.center) to (51.center);
		\draw [style=ThickLine, bend left=90] (52.center) to (53.center);
		\draw [style=ThickLine, bend right=90] (62.center) to (63.center);
		\draw [style=ThickLine, bend right=90] (64.center) to (65.center);
		\draw [style=ThickLine, bend left=90] (64.center) to (65.center);
		\draw [style=ThickLine, bend right=90] (66.center) to (67.center);
		\draw [style=ThickLine, bend right=90] (68.center) to (69.center);
		\draw [style=ThickLine, bend left=90] (68.center) to (69.center);
		\draw [style=ThickLine, bend right=90] (70.center) to (71.center);
		\draw [style=ThickLine, bend right=90] (72.center) to (73.center);
		\draw [style=ThickLine, bend left=90] (72.center) to (73.center);
		\draw [style=ThickLine, bend right=90] (74.center) to (75.center);
		\draw [style=ThickLine, bend right=90] (76.center) to (77.center);
		\draw [style=ThickLine, bend left=90] (76.center) to (77.center);
		\draw [style=ThickLine] (62.center) to (64.center);
		\draw [style=ThickLine] (65.center) to (63.center);
		\draw [style=ThickLine] (66.center) to (68.center);
		\draw [style=ThickLine] (69.center) to (67.center);
		\draw [style=ThickLine] (70.center) to (72.center);
		\draw [style=ThickLine] (71.center) to (73.center);
		\draw [style=ThickLine] (74.center) to (76.center);
		\draw [style=ThickLine] (75.center) to (77.center);
		\draw [style=DottedLine, bend left=90] (62.center) to (63.center);
		\draw [style=DottedLine, bend left=90] (66.center) to (67.center);
		\draw [style=DottedLine, bend left=90] (70.center) to (71.center);
		\draw [style=DottedLine, bend left=90] (74.center) to (75.center);
		\draw [style=ThickLine, bend right=90] (78.center) to (79.center);
		\draw [style=ThickLine, bend right=90] (80.center) to (81.center);
		\draw [style=ThickLine, bend right=90] (82.center) to (83.center);
		\draw [style=ThickLine, bend right=90] (84.center) to (85.center);
		\draw [style=DottedLine, bend left=90] (78.center) to (79.center);
		\draw [style=DottedLine, bend left=90] (80.center) to (81.center);
		\draw [style=DottedLine, bend left=90] (82.center) to (83.center);
		\draw [style=DottedLine, bend left=90] (84.center) to (85.center);
		\draw [style=ThickLine] (85.center) to (89.center);
		\draw [style=ThickLine] (89.center) to (84.center);
		\draw [style=ThickLine] (83.center) to (88.center);
		\draw [style=ThickLine] (88.center) to (82.center);
		\draw [style=ThickLine] (81.center) to (87.center);
		\draw [style=ThickLine] (87.center) to (80.center);
		\draw [style=ThickLine] (79.center) to (86.center);
		\draw [style=ThickLine] (86.center) to (78.center);
		\draw [style=ArrowLineRight] (94.center) to (95.center);
		\draw [style=ThickLine] (97.center) to (96.center);
		\draw [style=ThickLine, bend right=90, looseness=0.50] (103.center) to (104.center);
		\draw [style=ThickLine, bend right=90, looseness=0.50] (105.center) to (106.center);
		\draw [style=ThickLine, bend left=90, looseness=0.50] (105.center) to (106.center);
		\draw [style=ThickLine] (103.center) to (105.center);
		\draw [style=ThickLine] (106.center) to (104.center);
		\draw [style=DottedLine, bend left=90, looseness=0.50] (103.center) to (104.center);
		\draw [style=ThickLine, bend right=90, looseness=0.50] (107.center) to (108.center);
		\draw [style=DottedLine, bend left=90, looseness=0.50] (1.center) to (2.center);
		\draw [style=ThickLine, bend left=90, looseness=0.50] (107.center) to (108.center);
		\draw [style=ArrowLineRight] (113.center) to (114.center);
		\filldraw[fill=white, draw=white] (12.49, 9.25) ellipse (0.15 and 0.15);
		\filldraw[fill=white, draw=white] (17.2775, 9.3) ellipse (0.154 and 0.154);
		\filldraw[fill=white, draw=white] (17.2705, 6.695) ellipse (0.141 and 0.141);
		\filldraw[fill=white, draw=white] (12.49, 6.73) ellipse (0.15 and 0.15);
	\end{pgfonlayer}
\end{tikzpicture}
}
\caption{Depiction of the SymTFT / SymTh sliver in a single (i) and multi-throat (ii) string background.
In the multi-throat configuration, the tensor product of the physical boundary conditions
produces a pure state relative QFT and the bulk geometry results in a mixed state for topological
boundary conditions.}
\label{fig:SymTFTmultithroat}
\end{figure}
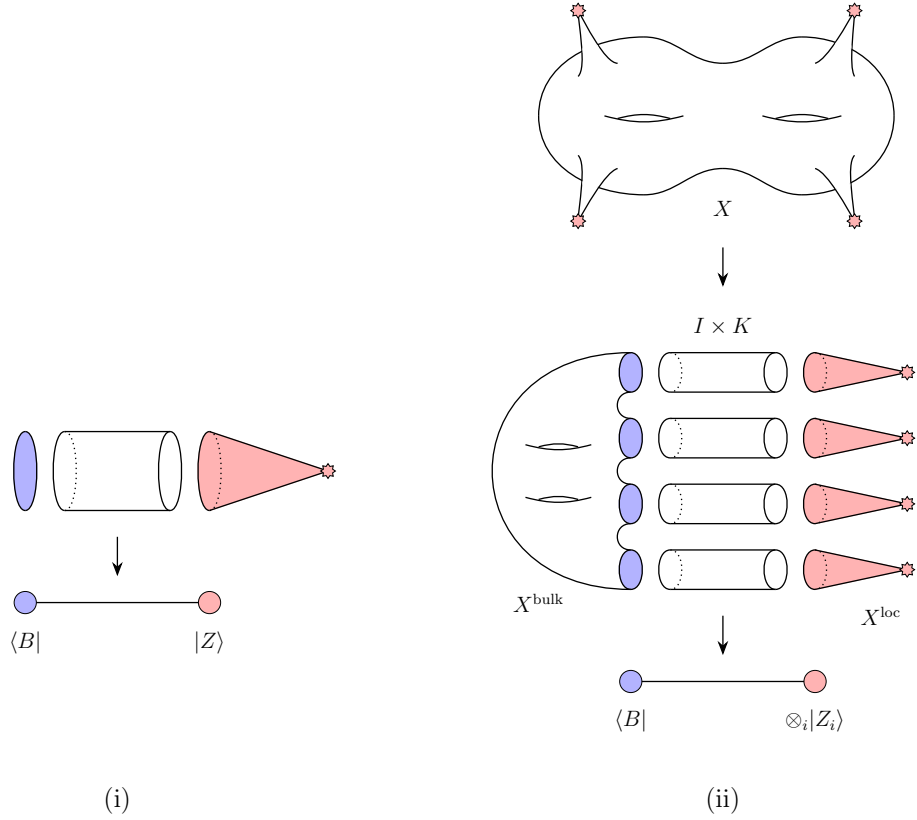

In the context of string compactifications, one often engineers a $D$-dimensional QFT
by starting with a higher-dimensional non-compact geometry with the degrees of freedom of the QFT localized at the tip of a
singular conical geometry. In actual string backgrounds, one can often entertain multiple conical regions, all glued together by a bulk geometry.
In this setting, when the bulk is large and $D$-dimensional gravity can be neglected, one can then speak of a more general notion of an entangled topological boundary condition for the whole multi-throat system. See figure \ref{fig:SymTFTmultithroat} for a depiction of a string compactification and its reduction to the SymTFT / SymTh.

To proceed, consider a string compactification with respect to an extra-dimensional space $X$ and split it into the disjoint union of all conical regions $X^{\text{loc}}$ as above and their complement $X^{\text{bulk}}=X\setminus X^\text{loc}$. In particular, we then have $X=X^{\text{loc}}\cup X^{\text{bulk}}$ and $X^{\text{loc}}=\sqcup_{i\:\!} X^{\text{loc}}_i$ where all $X^{\text{loc}}_i,X^{\text{bulk}}$ are connected subspaces of $X$ with boundary. We denote the common boundary
as $K=\partial X^{\text{loc}}=\partial X^{\text{bulk}}$ which has as many connected components as $X^{\text{loc}}$. Note that we take $X^{\text{bulk}}$ to be completely smooth and all singularities are contained in $X^\text{loc}$.
The above setup should more precisely be associated with the ``compressed" absolute system engineered by $X$. In order to decompress it and introduce the slab interval $I$ of the SymTFT / SymTh consider shrinking both $X^{\text{loc}}$ and $X^{\text{bulk}}$ using smooth deformation retractions. Abusing notation, we will continue to refer to these retracts as $X^{\text{loc}}$ and $X^{\text{bulk}}$ as they are homotopic. The space $X$ now decomposes naturally as
\begin{equation}
X=X^{\text{loc}} \cup \lb I\times K\rb \cup X^{\text{bulk}}\,.
\end{equation}
The SymTFT / SymTh is then supported on $I$ and is computed via compactification with respect to the compact cross-section $K$. After compactifying, each local model $X^{\text{loc}}_i$ engineers the SymTFT / SymTh state $|Z_i\rangle$ and to $X^{\text{loc}}$ we therefore associate overall the pure state
\begin{equation}
|Z\rangle \equiv \bigotimes_i |Z_i\rangle\,.
\end{equation}
Recall that $K$ has multiple connected components, and consequently, the SymTFT / SymTh constructed via compactification is a direct product of theories whose overall Hilbert space is simply the tensor product of the Hilbert spaces associated with the individual connected components of $K$. Completely analogously, $X^{\text{bulk}}$ is associated with some boundary condition which we denote $|B\rangle$, and the overall partition function of the system is $\langle B | Z \rangle$. When $X$ is compact, then $|B\rangle$ is usually a relative theory associated with supergravity bulk modes \cite{Cvetic:2023pgm}. When $X$ is non-compact, then $|B\rangle$ may be topological, but cases where these too support some relative theory associated with supergravity bulk modes can occur \cite{Baume:2023kkf}.

Consider now the case where the QFT has been decompressed onto the interval $I$ which supports a SymTFT, and expand the boundary condition $|B\rangle$ with respect to a basis of states
\begin{equation}
|B\rangle=\sum_{{i_1,\dots,i_m}} |B_{i_1,\dots,i_m}\rangle \langle B_{i_1,\dots,i_m}|B\rangle\equiv \sum_{{i_1,\dots,i_m}} b_{i_1,\dots,i_m}^* |B_{i_1,\dots,i_m}\rangle \,,
\end{equation}
where $m$ is the number of connected components of $K$ and $|B_{i_1,\dots,i_m}\rangle=|B_{i_1} \rangle\otimes \dots \otimes |B_{i_m} \rangle$ and the coefficients $b_{i_1,\dots,i_m} =\langle B_{i_1,\dots,i_m}|B\rangle^\dag$ are functions of the background fields. The overall partition function is thus:
\begin{equation}
\langle B | Z \rangle=\sum_{i_1,\dots,i_m}b_{i_1,\dots,i_m} \langle B_{i_1} | Z_{i_1}\rangle \dots\langle B_{i_m} | Z_{i_m} \rangle\,,
\end{equation}
which is an ensemble of direct products of absolute theories with partition functions $ \langle B_{i_k} | Z_{i_k}\rangle=Z_{i_k}(B_{i_k})$ weighted by $b_{i_1,\dots,i_m}$.

Finally, observe that once we trace over some of the $\vert B_{i} \rangle$ and $\vert Z_{j} \rangle$ states, we wind up with mixed states for the gapped / free and relative QFT boundaries. This tracing out is similar in spirit to that given in \cite{Balasubramanian:2020lux}, and it would be interesting to make the connection between these operations more precise.

\subsection{Holographic Systems}

Let us now turn to holographic examples featuring asymptotically AdS regions. In situations joined by wormhole configurations \cite{Witten:1999xp, Maldacena:2004rf}, the physical boundary conditions of the SymTFT / SymTh also need to be extended to cover entangled and mixed states. Specifying a single polarization from the start is physically cumbersome compared with using Wigner's function.

Recall that for a AdS/CFT\ pair with a single boundary, the SymTFT formalism can be viewed as defining
a \textquotedblleft topological sliver\textquotedblright\ of the
AdS/CFT\ correspondence \cite{Heckman:2024oot}.\footnote{See also
\cite{Aharony:1998qu, Witten:1998wy, Maldacena:2001ss}.} In more detail,
consider a large $N$ holographic CFT$_{D}$ on some $D$-dimensional manifold
$M_D$. Then, the SymTFT / SymTh is supported on the manifold $I\times M_{D}$, with $I$ an interval.
On one boundary, we have the relative quantum field theory, as specified by a physical
boundary condition $\left\vert Z \right\rangle $. On the other boundary, we have a topological / free boundary condition $\left\langle B \right\vert$ which specifies the polarization of the theory.

\begin{figure}
\centering
\scalebox{0.8}{
\begin{tikzpicture}
	\begin{pgfonlayer}{nodelayer}
		\node [style=none] (8) at (-1, 0) {};
		\node [style=none] (9) at (1, 0) {};
		\node [style=none] (10) at (0, 0) {};
		\node [style=none] (11) at (0, 0.5) {AdS/CFT};
		\node [style=none] (12) at (-3, 1.25) {};
		\node [style=none] (13) at (-3, -1.25) {};
		\node [style=none] (14) at (-6, 1.25) {};
		\node [style=none] (15) at (-6, -1.25) {};
		\node [style=none] (16) at (-3, 0) {$M_D$};
		\node [style=none] (17) at (-4.5, 1.75) {$I$};
		\node [style=none] (18) at (6, 1.25) {};
		\node [style=none] (19) at (6, -1.25) {};
		\node [style=none] (20) at (3, 1.25) {};
		\node [style=none] (21) at (3, -1.25) {};
		\node [style=none] (23) at (4.5, 1.75) {$I$};
		\node [style=none] (24) at (8.5, 0) {};
		\node [style=none] (25) at (7.5, 0) {AdS$_{D+1}$};
		\node [style=none] (27) at (-6, -1.75) {$\langle B |$};
		\node [style=none] (28) at (3, -1.75) {$\langle B |$};
		\node [style=none] (30) at (0, -2.75) {};
	\end{pgfonlayer}
	\begin{pgfonlayer}{edgelayer}
		\filldraw[fill=blue!30, draw=none] (-6,0) ellipse (0.54 and 1.25);
		\filldraw[fill=red!30, draw=none] (-3,0) ellipse (0.54 and 1.25);
		\filldraw[fill=red!30, draw=none] (6,0) ellipse (0.54 and 1.25);
		\filldraw[fill=blue!30, draw=none] (3,0) ellipse (0.54 and 1.25);
		\draw [style=ArrowLineRight] (10.center) to (9.center);
		\draw [style=ArrowLineRight] (10.center) to (8.center);
		\draw [style=ThickLine, bend left=90, looseness=0.75] (12.center) to (13.center);
		\draw [style=ThickLine, bend right=90, looseness=0.75] (14.center) to (15.center);
		\draw [style=ThickLine] (14.center) to (12.center);
		\draw [style=ThickLine] (15.center) to (13.center);
		\draw [style=ThickLine, bend left=90, looseness=0.75] (14.center) to (15.center);
		\draw [style=DottedLine, bend right=90, looseness=0.75] (12.center) to (13.center);
		\draw [style=ThickLine, bend left=90, looseness=0.75] (18.center) to (19.center);
		\draw [style=ThickLine, bend right=90, looseness=0.75] (20.center) to (21.center);
		\draw [style=ThickLine] (20.center) to (18.center);
		\draw [style=ThickLine] (21.center) to (19.center);
		\draw [style=ThickLine, bend left=90, looseness=0.75] (20.center) to (21.center);
		\draw [style=DottedLine, bend right=90, looseness=0.75] (18.center) to (19.center);
		\draw [style=ThickLine, in=90, out=0] (18.center) to (24.center);
		\draw [style=ThickLine, in=-90, out=0] (19.center) to (24.center);
	\end{pgfonlayer}
\end{tikzpicture}
}
\caption{Left: Sketch of the SymTFT / SymTh slab for a CFT with spacetime $M_D$. Right: The relative CFT is recast as its dual semi-classical gravitational theory on AdS$_{D+1}$ with boundary $M_D$. The SymTFT / SymTh is now naturally supported on an asymptotic sliver within the AdS$_{D+1}$ gravitational spacetime. }
\label{fig:AdS/CFT}
\end{figure}
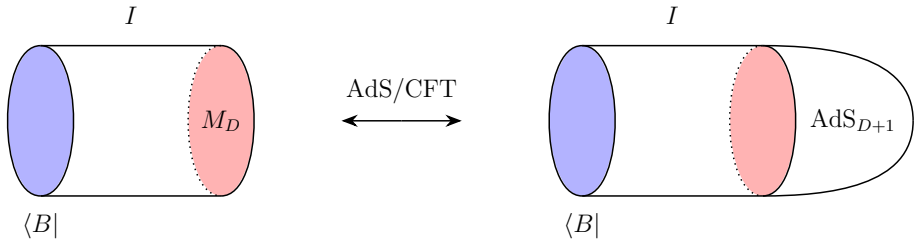

In the context of holographic
systems, we can view the physical boundary condition as equivalently being
captured by a gravitational theory on an asymptotically AdS$_{D+1}$ spacetime.
Then, the SymTFT / SymTh amounts to a small non-gravitational sliver which is
glued onto this asymptotic geometry \cite{Heckman:2024oot}. See figure \ref{fig:AdS/CFT} for a depiction.

Consider the well-studied case of a stack of $N$ D3-branes with semi-classical gravity dual AdS$_5 \times S^5$. Then $| Z \rangle$ is a relative $\mathfrak{su}_N$ gauge theory. The SymTFT was computed in \cite{Witten:1998wy} and the boundary condition $|B\rangle$ was discussed in \cite{Maldacena:2001ss}. The boundary condition $|B\rangle$ can be chosen to support a free $\mathfrak{u}_1$ gauge theory associated with the center of mass of the brane stack or taken to be topological / gapped.\footnote{Given the SymTFT action  $N/2\pi \int B_2dC_2$ the boundary condition $B_2=*C_2$  admits the center of mass degree of freedom to be added as a boundary mode and realizes the overall gauge algebra $\mathfrak{u}_N$, while Neumann / Dirichlet boundary conditions freeze the center of mass mode, and realize the overall gauge algebra $\mathfrak{su}_N$. As the bulk is topological, the center of mass of the brane stack is measured with respect to $|B\rangle$, and the distinguished cases correspond to an active or passive framing of coordinates. See \cite{Maldacena:2001ss, Heckman:2022xgu} for further details.} In the former case $\langle B |Z \rangle $ is the partition function of an absolute $U(N)$ gauge theory while in the latter the gauge group is $SU(N)/\mathbb{Z}_K$ with $K$ dividing $N$. In the former, the free abelian center of mass mode is coupled only topologically to the interacting degrees of freedom as specified by the SymTFT, they are edge modes to a common topological bulk.\footnote{One can also view the contribution from the center of mass degrees of freedom as adding a junction theory, in the sense of reference \cite{Baume:2023kkf}.}

\begin{figure}
\centering
\scalebox{0.8}{
\begin{tikzpicture}
	\begin{pgfonlayer}{nodelayer}
		\node [style=none] (0) at (-1.5, 2.25) {};
		\node [style=none] (1) at (-1.5, 1.5) {};
		\node [style=none] (2) at (-1.5, 1) {};
		\node [style=none] (3) at (-1.5, 0.25) {};
		\node [style=none] (4) at (-1.5, -0.25) {};
		\node [style=none] (5) at (-1.5, -1) {};
		\node [style=none] (6) at (-1.5, -1.5) {};
		\node [style=none] (7) at (-1.5, -2.25) {};
		\node [style=none] (8) at (-3.25, 0) {};
		\node [style=none] (9) at (-3, -0.04) {};
		\node [style=none] (10) at (-2.5, -0.04) {};
		\node [style=none] (11) at (-2.25, 0) {};
		\node [style=none] (16) at (-1, 2.25) {};
		\node [style=none] (17) at (-1, 1.5) {};
		\node [style=none] (18) at (1, 2.25) {};
		\node [style=none] (19) at (1, 1.5) {};
		\node [style=none] (20) at (-1, 1) {};
		\node [style=none] (21) at (-1, 0.25) {};
		\node [style=none] (22) at (1, 1) {};
		\node [style=none] (23) at (1, 0.25) {};
		\node [style=none] (24) at (-1, -0.25) {};
		\node [style=none] (25) at (-1, -1) {};
		\node [style=none] (26) at (1, -0.25) {};
		\node [style=none] (27) at (1, -1) {};
		\node [style=none] (28) at (-1, -1.5) {};
		\node [style=none] (29) at (-1, -2.25) {};
		\node [style=none] (30) at (1, -1.5) {};
		\node [style=none] (31) at (1, -2.25) {};
		\node [style=none] (32) at (1.5, 2.25) {};
		\node [style=none] (33) at (1.5, 1.5) {};
		\node [style=none] (34) at (1.5, 1) {};
		\node [style=none] (35) at (1.5, 0.25) {};
		\node [style=none] (36) at (1.5, -0.25) {};
		\node [style=none] (37) at (1.5, -1) {};
		\node [style=none] (38) at (1.5, -1.5) {};
		\node [style=none] (39) at (1.5, -2.25) {};
		\node [style=none] (40) at (2.25, 0) {};
		\node [style=none] (41) at (2.5, -0.04) {};
		\node [style=none] (42) at (3, -0.04) {};
		\node [style=none] (43) at (3.25, 0) {};
		\node [style=none] (48) at (-0.75, 1.875) {};
		\node [style=none] (49) at (0.75, 1.875) {};
		\node [style=none] (50) at (1.75, 1.875) {};
		\node [style=none] (51) at (-1.75, 1.875) {};
		\node [style=none] (52) at (-1.75, -1.875) {};
		\node [style=none] (53) at (1.75, -1.875) {};
		\node [style=none] (54) at (-0.75, -1.875) {};
		\node [style=none] (55) at (0.75, -1.875) {};
		\node [style=none] (56) at (0.2, 1.875) {};
		\node [style=none] (57) at (0.35, 1.875) {};
		\node [style=none] (58) at (0.175, 2) {};
		\node [style=none] (59) at (0.175, 1.75) {};
		\node [style=none] (60) at (0.25, 2.15) {};
		\node [style=none] (61) at (0.25, 1.60) {};
		\node [style=none] (62) at (5, 0) {$P_{D+1}$};
		\node [style=none] (63) at (-5, 0) {$T_{D+1}$};
		\node [style=none] (64) at (1.5, 2.75) {$|Z_{i_1}\rangle$};
		\node [style=none] (65) at (-1.5, 2.75) {$\langle B_{i_1}|$};
		\node [style=none] (66) at (1.5, -2.75) {$|Z_{i_m}\rangle$};
		%\node [style=none] (67) at (2, -0.5) {$...$};
		%\node [style=none] (68) at (2, 0.75) {$...$};
		\node [style=none] (69) at (-1.5, -2.75) {$\langle B_{i_m}|$};
		%\node [style=none] (70) at (-2, -0.5) {$...$};
		%\node [style=none] (71) at (-2, 0.75) {$...$};
		\node [style=none] (72) at (0, -3.25) {};
		\node [style=none] (73) at (0, 2.75) {$\epsilon$};
	\end{pgfonlayer}
	\begin{pgfonlayer}{edgelayer}
		\filldraw[fill=red!30, draw=none] (1.5, 0.625) ellipse (0.175 and 0.375);
		\filldraw[fill=red!30, draw=none] (1.5, 1.875) ellipse (0.175 and 0.375);
		\filldraw[fill=red!30, draw=none] (1.5, -0.625) ellipse (0.175 and 0.375);
		\filldraw[fill=red!30, draw=none] (1.5, -1.875) ellipse (0.175 and 0.375);
		\filldraw[fill=blue!30, draw=none] (-1.5, 0.625) ellipse (0.175 and 0.375);
		\filldraw[fill=blue!30, draw=none] (-1.5, 1.875) ellipse (0.175 and 0.375);
		\filldraw[fill=blue!30, draw=none] (-1.5, -0.625) ellipse (0.175 and 0.375);
		\filldraw[fill=blue!30, draw=none] (-1.5, -1.875) ellipse (0.175 and 0.375);
		\draw [style=ThickLine, bend right=90, looseness=1.75] (1.center) to (2.center);
		\draw [style=ThickLine, in=180, out=-180, looseness=1.75] (3.center) to (4.center);
		\draw [style=ThickLine, bend right=90, looseness=1.75] (5.center) to (6.center);
		\draw [style=ThickLine, bend right=90, looseness=2.00] (0.center) to (7.center);
		\draw [style=ThickLine, bend left=15] (9.center) to (10.center);
		\draw [style=ThickLine, bend left=15] (11.center) to (8.center);
		\draw [style=ThickLine, bend right=90, looseness=0.75] (0.center) to (1.center);
		\draw [style=ThickLine, bend right=270, looseness=0.75] (3.center) to (2.center);
		\draw [style=ThickLine, bend right=90, looseness=0.75] (4.center) to (5.center);
		\draw [style=ThickLine, bend right=90, looseness=0.75] (6.center) to (7.center);
		\draw [style=ThickLine, bend left=90, looseness=0.75] (0.center) to (1.center);
		\draw [style=ThickLine, bend left=90, looseness=0.75] (2.center) to (3.center);
		\draw [style=ThickLine, bend left=90, looseness=0.75] (4.center) to (5.center);
		\draw [style=ThickLine, bend left=90, looseness=0.75] (6.center) to (7.center);
		\draw [style=ThickLine, bend right=90, looseness=0.75] (16.center) to (17.center);
		\draw [style=ThickLine, bend right=90, looseness=0.75] (18.center) to (19.center);
		\draw [style=ThickLine, bend left=90, looseness=0.75] (18.center) to (19.center);
		\draw [style=ThickLine, bend right=90, looseness=0.75] (20.center) to (21.center);
		\draw [style=ThickLine, bend right=90, looseness=0.75] (22.center) to (23.center);
		\draw [style=ThickLine, bend left=90, looseness=0.75] (22.center) to (23.center);
		\draw [style=ThickLine, bend right=90, looseness=0.75] (24.center) to (25.center);
		\draw [style=ThickLine, bend right=90, looseness=0.75] (26.center) to (27.center);
		\draw [style=ThickLine, bend left=90, looseness=0.75] (26.center) to (27.center);
		\draw [style=ThickLine, bend right=90, looseness=0.75] (28.center) to (29.center);
		\draw [style=ThickLine, bend right=90, looseness=0.75] (30.center) to (31.center);
		\draw [style=ThickLine, bend left=90, looseness=0.75] (30.center) to (31.center);
		\draw [style=ThickLine] (16.center) to (18.center);
		\draw [style=ThickLine] (19.center) to (17.center);
		\draw [style=ThickLine] (20.center) to (22.center);
		\draw [style=ThickLine] (23.center) to (21.center);
		\draw [style=ThickLine] (24.center) to (26.center);
		\draw [style=ThickLine] (25.center) to (27.center);
		\draw [style=ThickLine] (28.center) to (30.center);
		\draw [style=ThickLine] (29.center) to (31.center);
		\draw [style=DottedLine, bend left=90, looseness=0.75] (16.center) to (17.center);
		\draw [style=DottedLine, bend left=90, looseness=0.75] (20.center) to (21.center);
		\draw [style=DottedLine, bend left=90, looseness=0.75] (24.center) to (25.center);
		\draw [style=DottedLine, bend left=90, looseness=0.75] (28.center) to (29.center);
		\draw [style=ThickLine, bend left=90, looseness=1.75] (33.center) to (34.center);
		\draw [style=ThickLine, bend left=90, looseness=1.75] (35.center) to (36.center);
		\draw [style=ThickLine, bend left=90, looseness=1.75] (37.center) to (38.center);
		\draw [style=ThickLine, bend right=90, looseness=0.75] (32.center) to (33.center);
		\draw [style=ThickLine, bend left=90, looseness=0.75] (35.center) to (34.center);
		\draw [style=ThickLine, bend right=105, looseness=0.75] (36.center) to (37.center);
		\draw [style=ThickLine, bend right=90, looseness=0.75] (38.center) to (39.center);
		\draw [style=ThickLine, bend left=90, looseness=2.00] (32.center) to (39.center);
		\draw [style=ThickLine, bend left=15] (41.center) to (42.center);
		\draw [style=ThickLine, bend left=15] (43.center) to (40.center);
		\draw [style=DottedLine, bend left=90, looseness=0.75] (32.center) to (33.center);
		\draw [style=DottedLine, bend left=90, looseness=0.75] (34.center) to (35.center);
		\draw [style=DottedLine, bend left=90, looseness=0.75] (36.center) to (37.center);
		\draw [style=DottedLine, bend left=90, looseness=0.75] (38.center) to (39.center);
		\draw [style=PurpleLine, snake it, bend left=90, looseness=1.7] (50.center) to (53.center);
		\draw [style=PurpleLine, snake it] (55.center) to (54.center);
		\draw [style=PurpleLine, snake it, bend left=90, looseness=1.7] (52.center) to (51.center);
		\draw [style=PurpleLine, snake it] (48.center) to (56.center);
		\draw [style=PurpleLine, snake it] (57.center) to (49.center);
		\draw [style=BrownLine, in=-180, out=90, looseness=0.75] (58.center) to (60.center);
		\draw [style=BrownLine, in=-180, out=-90, looseness=0.75] (59.center) to (61.center);
		\draw [style=BrownLine, in=0, out=0, looseness=0.25] (61.center) to (60.center);
	\end{pgfonlayer}
\end{tikzpicture}
}
\caption{Depiction of a more general entangled state in the SymTFT / SymTh formalism, embedded in a holographic system. In this setting, preparation of the general state involves a gravitational path integral over the physical region $P_{D+1}$ and a topological / free path integral over the region $T_{D+1}$. We also depict a defect (purple) threading the geometry and its linking with a symmetry operator (brown) in the $\epsilon$-sliver.}
\label{fig:MultiSymTFTsliver}
\end{figure}
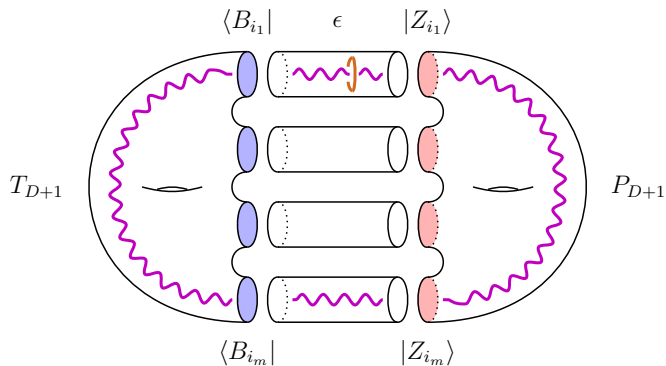

We now generalize this to situations with more than one boundary. To this
end, consider starting with a general $(D+1)$-dimensional manifold $M_{D+1}$ with no boundaries.
We cut this into two pieces with the same boundary, namely
$P_{D+1}$ and $T_{D+1}$ such that $\partial P_{D+1} = \partial\overline{T_{D+1}} = M_{D}$.
In general, $M_{D}$ might consist of several components:%
\begin{equation}
M_{D}=\underset{i}{{\displaystyle\bigsqcup}}\,M_{D}^{(i)} \, .
\end{equation}
We specify a gravitational theory (with prescribed boundary conditions) on
$P_{D+1}$, and a SymTFT\ / SymTh (with prescribed boundary conditions) on
$T_{D+1}$. From $P_{D+1}$ and $T_{D+1}$ we get pure states:%
\begin{align}
\left\vert Z\right\rangle  &  \equiv\underset{i_{1},...,i_{m}}{\sum}%
z_{i_{1},...,i_{m}}\left\vert Z_{i_{1}}\right\rangle \otimes...\otimes\left\vert
Z_{i_{m}}\right\rangle \\
\left\vert B\right\rangle  &  \equiv\underset{i_{1},...,i_{m}}{\sum}%
b_{i_{1},...,i_{m}}\left\vert B{i_{1}}\right\rangle \otimes...\otimes\left\vert
B_{i_{m}}\right\rangle ,
\end{align}
and a corresponding overlap via $\left\langle
B|Z\right\rangle =Z(B)$ (see figure \ref{fig:MultiSymTFTsliver}). Abstractly, $T_{D+1}$ can simply be viewed as a geometrization of the boundary condition, the SymTFT assigns a state to $T_{D+1}$ that serves as a boundary condition. In top-down constructions, we can also think of the above setting as arising from, say, a compact geometry with multiple brane stacks dispersed. After taking near-horizon limits centered on each stack, multiple AdS throats develop, and after possibly tracing out some or all degrees of freedom, we find the bulk beyond the throats to specify gapped / free boundary conditions.

The state $\left\vert Z\right\rangle $
consists of a linear combination of tensor products of different relative
theories. In the context of theories with a bulk gravity dual, we interpret
this as a Euclidean wormhole connecting different large $N$ theories. The
state $\left\vert B\right\rangle $ consists of a linear combination of
different topological / free boundary conditions. In the context of a string
compactification, we interpret this as taking a local model with prescribed
boundary conditions and gluing it to other local models via some bulk
extra-dimensional geometry.

We can also fix our attention on just one of the factors in the tensor
product. One way to do this is to perform a partial trace over the other
sectors. In general, this results in a mixed state of relative theories and
topological boundary conditions. In particular, this makes it somewhat awkward
to specify a single absolute quantum field theory. It is therefore somewhat
more natural to use Wigner's function so that we can deal with all possible
choices of polarizations simultaneously.

With this in mind, we now consider Wigner's function for the physical state
$\left\vert Z\right\rangle $ as specified above. Then, returning to
(\ref{eq:WignerMixed}), we have:%
\begin{equation}\label{eq:NoPolarization2}
W^{[r,s]}_{Z}(\mathbb{B})=\left\langle Z\right\vert \mathbb{U}^{[s]}_{\mathbb{B}}\:\!\mathcal{C}\:\!%
\mathbb{U}_{\mathbb{B}}^{[r] \, \dag}\left\vert Z\right\rangle ,
\end{equation}
where here, the operator $ \mathbb{U}^{[s]}_{\mathbb{B}}\:\!\mathcal{C}\:\!%
\mathbb{U}_{\mathbb{B}}^{[r] \, \dag}$ is defined intrinsically in the SymTFT$_{D+1}$
region of the geometry. As such, we can simultaneously handle all possible
polarizations. Geometrically, we also get a linking that extends over all of
$M_{D+1}$, see figure \ref{fig:MultiSymTFTsliver}.

\subsubsection{AdS-Schwarzschild Geometry}

As a specific holographic example, we now turn to the eternal
black hole in AdS \cite{Israel:1976ur, Maldacena:2001kr}. In Euclidean
signature, this background has a single boundary, but in Lorentzian signature,
we have two boundaries. To leading order, one can interpret this bulk system
as the gravity dual of two entangled Lorentzian signature CFTs, which we
assume to be on manifolds of the form $\mathbb{R}_{\text{time}}\times
S^{D-1}=M_{D}$. In this setting, the \textquotedblleft partition
function\textquotedblright\ just amounts to evaluating the path integral with
prescribed boundary conditions in the past and future.

There are now two natural notions of time. First, there is $t$, the time of
each boundary CFT. Second, we have $r$, the radial coordinate of the associated
symmetry theory / gravity dual. To begin with, we shall mainly focus on the
radial direction as the time coordinate, but return to the other notion of time evolution later.

Observe that we now have a two-sided system, and as such, we also get two
SymTFTs / SymThs. Including the gravitational bulk, we observe that there is a natural path integral
with left and right boundary conditions $B_L$ and $B_R$:
\begin{equation}
Z_{\mathrm{grav}}(B_L, B_R) = \int_{B_L}^{B_R} d \Phi \, e^{i S[\Phi]},
\end{equation}
where we leave implicit the boundary conditions in the past and future.

Let us now interpret this expression in the SymTFT framework. Along these lines, a general comment here is that the ``state'' we are discussing is that coming from the SymTFT / SymTh formalism, i.e., it is a state of a $(D+1)$-dimensional system. This is to be contrasted with the states one would discuss of the corresponding entangled boundary CFTs. Indeed, while one views AdS-Schwarzschild as the purification of the thermofield double state of the CFT$_{D}$, that is a statement about a lower-dimensional system. In the present setting, building a linear combination of states is more akin to engineering an ensemble of different theories. Of course, changing the notion of time and thus the construction of states relates the two pictures via the path integral.

With this in mind, introduce the density matrices associated with SymTFT states:
\begin{equation}
\widehat{\rho}_{Z}=\underset{i}{\sum}z_{i}\left\vert Z_{i}\right\rangle \left\langle
Z_{i}\right\vert \text{ \ \ and \ \ }\widehat{\rho}_{B}=\underset{j}{\sum}%
b_{j}\left\vert B_{j}\right\rangle \left\langle B_{j}\right\vert .
\end{equation}
Here, we have chosen to work in the respective eigenbases so that the density matrices are diagonal.
By inspection, we can equivalently write the gravitational path integral as
the expectation value:
\begin{equation}
Z_{\text{grav}}(p_B)=\text{Tr}\left(  \widehat{\rho}_{B} \widehat{\rho}_{Z}\right) = \underset{i,j}{\sum}Z_{i}(B_{j})M_{ij}\overline{Z_{i}%
(B_{j})},
\end{equation}
with $M_{ij} = z_i b_j$. We can also write this as the expectation value obtained
from Wigner's function:
\begin{equation}
Z_{\text{grav}}(p_B)=\underset{\mathbb{B}}{\sum}W^{[r,s]}_{Z}(\mathbb{B})p_{B}%
(\mathbb{B}),
\end{equation}
where $p_{B}(\mathbb{B})$ is the classical function corresponding to the density matrix operator $\widehat{\rho}_{B}$. Here, Wigner's function is given by:
\begin{equation}
W_{Z}^{[r,s]}(\mathbb{B})=\text{Tr}(\widehat{\rho}_{Z}\mathbb{U}^{[s]}_{\mathbb{B}}\:\!\mathcal{C}\:\!%
\mathbb{U}_{\mathbb{B}}^{[r] \, \dag}),
\end{equation}
in the obvious notation.

\subsubsection{Gluing to the Thermofield Double State}

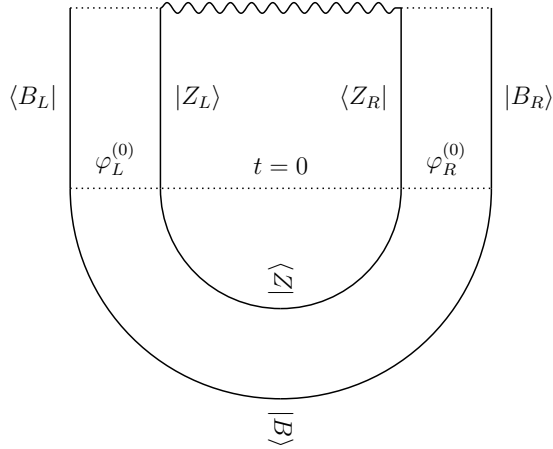
\begin{figure}
\centering
\scalebox{0.8}{
\begin{tikzpicture}
	\begin{pgfonlayer}{nodelayer}
		\node [style=none] (0) at (-2, 2) {};
		\node [style=none] (1) at (2, 2) {};
		\node [style=none] (2) at (0, 0) {};
		\node [style=none] (3) at (-3.5, 2) {};
		\node [style=none] (4) at (3.5, 2) {};
		\node [style=none] (5) at (0, -1.5) {};
		\node [style=none] (6) at (-2, 5) {};
		\node [style=none] (7) at (2, 5) {};
		\node [style=none] (8) at (-3.5, 5) {};
		\node [style=none] (9) at (3.5, 5) {};
		\node [style=none] (10) at (-4.125, 3.5) {$ \langle B_L |$};
		\node [style=none] (11) at (-1.375, 3.5) {$| Z_L\rangle$};
		\node [style=none] (12) at (1.375, 3.5) {$\langle Z_R |$};
		\node [style=none] (13) at (4.125, 3.5) {$|B_R\rangle$};
		\node [style=none] (14) at (0, 2.375) {$t=0$};
		\node [style=none] (15) at (0, -2) {\rotatebox{-90}{$| B\:\! \rangle $}};
		\node [style=none] (16) at (0, 0.5) {\rotatebox{-90}{$\langle Z |$}};
		\node [style=none] (17) at (-2.75, 2.5) {$\varphi_L^{(0)}$};
		\node [style=none] (18) at (2.75, 2.5) {$\varphi_R^{(0)}$};
	\end{pgfonlayer}
	\begin{pgfonlayer}{edgelayer}
		\draw [style=ThickLine, in=180, out=-90] (0.center) to (2.center);
		\draw [style=ThickLine, in=0, out=-90] (1.center) to (2.center);
		\draw [style=ThickLine, in=0, out=-90] (4.center) to (5.center);
		\draw [style=ThickLine, in=-90, out=180] (5.center) to (3.center);
		\draw [style=DottedLine] (3.center) to (4.center);
		\draw [style=ThickLine] (8.center) to (3.center);
		\draw [style=ThickLine] (6.center) to (0.center);
		\draw [style=ThickLine] (7.center) to (1.center);
		\draw [style=ThickLine] (9.center) to (4.center);
		\draw [style=DottedLine] (8.center) to (6.center);
		\draw [style=DottedLine] (7.center) to (9.center);
		\draw [style=ThickLine, snake it] (6.center) to (7.center);
	\end{pgfonlayer}
\end{tikzpicture}}
\caption{Depiction of the preparation of the thermofield double state, its gravity dual, and the gluing to a SymTFT / SymTh sliver. Below the $t = 0$ dashed line, we have a single-sided gravity dual in Euclidean signature, which prepares the thermofield double. This is then glued to the two-sided SymTFT / SymTh geometry in Lorentzian signature (above $t = 0$). There are corresponding relative QFT and topological / free boundary conditions, as well as field configurations at $t = 0$ which are suitably glued together by the path integral defined in each region.}
\label{fig:AdSchw}
\end{figure}

It is also of interest to now specify fixed boundary conditions at $t = 0$ as specified by the thermofield double
state of a CFT at inverse temperature $\beta$. Recall from reference
\cite{Maldacena:2001kr} that the AdS-Schwarzschild geometry can also be viewed
as the CFT thermofield double state%
\begin{equation}
\left\vert \Psi(t=0)\right\rangle =\frac{1}{\sqrt{Z_{\beta}}}\underset{E_{L}%
=E_{R}=E}{\sum}e^{-\beta E/2}\left\vert E_{L},E_{R}\right\rangle ,
\end{equation}
namely we work at a fixed time slice $(t=0)$ and prepare a corresponding state
via the Euclidean path integral on the background $I_{\beta/2}\times S^{D-1}$
\cite{Maldacena:2001kr}. The SymTFT / SymTh for this setting consists of a single
physical boundary condition and a single topological boundary condition.
Additionally, we have to specify the boundary conditions for the path integral
at the two ends of the interval, which we label as $\varphi_{L}^{(0)}$ and
$\varphi_{R}^{(0)}$. We glue the left and right ends of this interval to the
Lorentzian signature partition functions. Taking all of this into account, the wavefunction
depends on a single topological boundary condition and two gluing conditions
on the left and right. We write all of this as: $Z_{\text{Euc}}(B;\varphi
_{L}^{(0)},\varphi_{R}^{(0)})$, which is prepared via the Euclidean path
integral on $I\times\left(  I_{\beta/2}\times S^{D-1}\right)  $.

Turning next to the Lorentzian signature path integrals, we now have a single
fixed topological boundary condition $B=B_{L}=B_{R}$.
In this patch, we have $Z_{\text{Lor}}(B_{L}=B,B_{R}=B,\varphi_{L}%
^{(0)},\varphi_{R}^{(0)})$, in the obvious notation. Putting everything
together, we get:%
\begin{equation}
Z_{\text{grav}}(B_L,B_R) = 
\begin{cases}
\int d\varphi_{L}^{(0)}d\varphi_{R}^{(0)} \, Z_{\text{Lor}}(B,B,\varphi_{L}^{(0)},\varphi_{R}^{(0)})Z_{\text{Euc}}(B;\varphi_{L}^{(0)},\varphi_{R}^{(0)}) & B_L = B_R = B \\
0 & B_L \neq B_R
\end{cases}
.
\end{equation}
Here, the answer vanishes if $B_L \neq B_R$ because it is impossible to glue these topological boundary conditions to the Euclidean section.
Observe that the topological boundary conditions on the two sides of the eternal black hole are now entangled, and this is a consequence of the geometrical connection between these topological boundaries.
See figure \ref{fig:AdSchw} for a depiction. 
Thus, our construction provides an explicit example of the ER = EPR proposal \cite{Maldacena:2013xja} in the context of SymTFT/SymTh, wherein an ER bridge connecting the topological boundaries results in the states on these boundaries being entangled.

\newpage

\section*{Acknowledgements}

We thank V. Chakrabhavi, C. Cummings, M. Del Zotto, E. Torres, and X. Yu for helpful discussions.
%JJH thanks KITP for denying his application to the program on generalized symmetries in quantum field theory.
The work of JJH is supported in part by a University Research Foundation grant
at the University of Pennsylvania as well as by BSF grant 2022100.
The work of JJH and CM is supported by DOE (HEP) Award DE-SC0013528.
The work of MH is supported by the Marie Skłodowska-Curie Actions under the European Union’s
Horizon 2020 research and innovation programme, grant agreement number \#101109804.
MH acknowledges support from the the VR Centre for Geometry and Physics (VR grant
No. 2022-06593). The work of CM is also supported by the DOE through QuantISED grant DE-SC0020360.

\bibliographystyle{utphys}
\bibliography{WignerSymTFT}

\providecommand{\href}[2]{#2}\begingroup\raggedright\begin{thebibliography}{10}

\bibitem{Gaiotto:2014kfa}
D.~Gaiotto, A.~Kapustin, N.~Seiberg, and B.~Willett, ``{Generalized Global
  Symmetries},'' \href{http://dx.doi.org/10.1007/JHEP02(2015)172}{{\em JHEP}
  {\bfseries 02} (2015) 172}, \href{http://arxiv.org/abs/1412.5148}{{\ttfamily
  arXiv:1412.5148 [hep-th]}}.

\bibitem{Reshetikhin:1991tc}
N.~Reshetikhin and V.~G. Turaev, ``{Invariants of three manifolds via link
  polynomials and quantum groups},''
  \href{http://dx.doi.org/10.1007/BF01239527}{{\em Invent. Math.} {\bfseries
  103} (1991) 547--597}.

\bibitem{TURAEV1992865}
V.~Turaev and O.~Viro, ``State sum invariants of 3-manifolds and quantum
  6j-symbols,'' {\em Topology} {\bfseries 31} no.~4, (1992) 865--902.

\bibitem{Barrett:1993ab}
J.~W. Barrett and B.~W. Westbury, ``{Invariants of piecewise linear three
  manifolds},'' \href{http://dx.doi.org/10.1090/S0002-9947-96-01660-1}{{\em
  Trans. Am. Math. Soc.} {\bfseries 348} (1996) 3997--4022},
  \href{http://arxiv.org/abs/hep-th/9311155}{{\ttfamily arXiv:hep-th/9311155}}.

\bibitem{Witten:1998wy}
E.~Witten, ``{AdS / CFT correspondence and topological field theory},''
  \href{http://dx.doi.org/10.1088/1126-6708/1998/12/012}{{\em JHEP} {\bfseries
  12} (1998) 012}, \href{http://arxiv.org/abs/hep-th/9812012}{{\ttfamily
  arXiv:hep-th/9812012}}.

\bibitem{Fuchs:2002cm}
J.~Fuchs, I.~Runkel, and C.~Schweigert, ``{TFT construction of RCFT correlators
  1. Partition functions},''
  \href{http://dx.doi.org/10.1016/S0550-3213(02)00744-7}{{\em Nucl. Phys. B}
  {\bfseries 646} (2002) 353--497},
  \href{http://arxiv.org/abs/hep-th/0204148}{{\ttfamily arXiv:hep-th/0204148}}.

\bibitem{Kirillov2010TuraevViroIA}
A.~K. Jr. and B.~Balsam, ``Turaev-viro invariants as an extended tqft,''
  \href{http://arxiv.org/abs/1004.1533}{{\ttfamily arXiv:1004.1533 [math.GT]}}.

\bibitem{Kapustin:2010if}
A.~Kapustin and N.~Saulina, ``{Surface operators in 3d Topological Field Theory
  and 2d Rational Conformal Field Theory},''
  \href{http://arxiv.org/abs/1012.0911}{{\ttfamily arXiv:1012.0911 [hep-th]}}.

\bibitem{Kitaev2011ModelsFG}
A.~Y. Kitaev and L.~Kong, ``Models for gapped boundaries and domain walls,''
  {\em Communications in Mathematical Physics} {\bfseries 313} (2011) 351--373.

\bibitem{Fuchs:2012dt}
J.~Fuchs, C.~Schweigert, and A.~Valentino, ``{Bicategories for boundary
  conditions and for surface defects in 3-d TFT},''
  \href{http://dx.doi.org/10.1007/s00220-013-1723-0}{{\em Commun. Math. Phys.}
  {\bfseries 321} (2013) 543--575},
  \href{http://arxiv.org/abs/1203.4568}{{\ttfamily arXiv:1203.4568 [hep-th]}}.

\bibitem{Freed:2012bs}
D.~S. Freed and C.~Teleman, ``{Relative quantum field theory},''
  \href{http://dx.doi.org/10.1007/s00220-013-1880-1}{{\em Commun. Math. Phys.}
  {\bfseries 326} (2014) 459--476},
  \href{http://arxiv.org/abs/1212.1692}{{\ttfamily arXiv:1212.1692 [hep-th]}}.

\bibitem{Heckman:2017uxe}
J.~J. Heckman and L.~Tizzano, ``{6D Fractional Quantum Hall Effect},''
  \href{http://dx.doi.org/10.1007/JHEP05(2018)120}{{\em JHEP} {\bfseries 05}
  (2018) 120}, \href{http://arxiv.org/abs/1708.02250}{{\ttfamily
  arXiv:1708.02250 [hep-th]}}.

\bibitem{Freed:2018cec}
D.~S. Freed and C.~Teleman, ``{Topological dualities in the Ising model},''
  \href{http://dx.doi.org/10.2140/gt.2022.26.1907}{{\em Geom. Topol.}
  {\bfseries 26} (2022) 1907--1984},
  \href{http://arxiv.org/abs/1806.00008}{{\ttfamily arXiv:1806.00008
  [math.AT]}}.

\bibitem{Gaiotto:2020iye}
D.~Gaiotto and J.~Kulp, ``{Orbifold groupoids},''
  \href{http://dx.doi.org/10.1007/JHEP02(2021)132}{{\em JHEP} {\bfseries 02}
  (2021) 132}, \href{http://arxiv.org/abs/2008.05960}{{\ttfamily
  arXiv:2008.05960 [hep-th]}}.

\bibitem{Apruzzi:2021nmk}
F.~Apruzzi, F.~Bonetti, I.~Garcia~Etxebarria, S.~S. Hosseini, and
  S.~Schafer-Nameki, ``{Symmetry TFTs from String Theory},''
  \href{http://dx.doi.org/10.1007/s00220-023-04737-2}{{\em Commun. Math. Phys.}
  {\bfseries 402} no.~1, (2023) 895--949},
  \href{http://arxiv.org/abs/2112.02092}{{\ttfamily arXiv:2112.02092
  [hep-th]}}.

\bibitem{Freed:2022qnc}
D.~S. Freed, G.~W. Moore, and C.~Teleman, ``{Topological symmetry in quantum
  field theory},'' \href{http://arxiv.org/abs/2209.07471}{{\ttfamily
  arXiv:2209.07471 [hep-th]}}.

\bibitem{Kaidi:2022cpf}
J.~Kaidi, K.~Ohmori, and Y.~Zheng, ``{Symmetry TFTs for Non-invertible
  Defects},'' \href{http://dx.doi.org/10.1007/s00220-023-04859-7}{{\em Commun.
  Math. Phys.} {\bfseries 404} no.~2, (2023) 1021--1124},
  \href{http://arxiv.org/abs/2209.11062}{{\ttfamily arXiv:2209.11062
  [hep-th]}}.

\bibitem{Baume:2023kkf}
F.~Baume, J.~J. Heckman, M.~H\"ubner, E.~Torres, A.~P. Turner, and X.~Yu,
  ``{SymTrees and Multi-Sector QFTs},''
  \href{http://dx.doi.org/10.1103/PhysRevD.109.106013}{{\em Phys. Rev. D}
  {\bfseries 109} no.~10, (2024) 106013},
  \href{http://arxiv.org/abs/2310.12980}{{\ttfamily arXiv:2310.12980
  [hep-th]}}.

\bibitem{Brennan:2024fgj}
T.~D. Brennan and Z.~Sun, ``{A SymTFT for continuous symmetries},''
  \href{http://dx.doi.org/10.1007/JHEP12(2024)100}{{\em JHEP} {\bfseries 12}
  (2024) 100}, \href{http://arxiv.org/abs/2401.06128}{{\ttfamily
  arXiv:2401.06128 [hep-th]}}.

\bibitem{Heckman:2024oot}
J.~J. Heckman, M.~H\"ubner, and C.~Murdia, ``{On the Holographic Dual of a
  Topological Symmetry Operator},''
  \href{http://dx.doi.org/10.1103/PhysRevD.110.046007}{{\em Phys. Rev. D}
  {\bfseries 110} no.~4, (2024) 046007},
  \href{http://arxiv.org/abs/2401.09538}{{\ttfamily arXiv:2401.09538
  [hep-th]}}.

\bibitem{Argurio:2024oym}
R.~Argurio, F.~Benini, M.~Bertolini, G.~Galati, and P.~Niro, ``{On the symmetry
  TFT of Yang-Mills-Chern-Simons theory},''
  \href{http://dx.doi.org/10.1007/JHEP07(2024)130}{{\em JHEP} {\bfseries 07}
  (2024) 130}, \href{http://arxiv.org/abs/2404.06601}{{\ttfamily
  arXiv:2404.06601 [hep-th]}}.

\bibitem{Heckman:2024zdo}
J.~J. Heckman and M.~H\"ubner, ``{Celestial Topology, Symmetry Theories, and
  Evidence for a Non-SUSY D3-Brane CFT},''
  \href{http://arxiv.org/abs/2406.08485}{{\ttfamily arXiv:2406.08485
  [hep-th]}}.

\bibitem{Cvetic:2024dzu}
M.~Cveti\v{c}, R.~Donagi, J.~J. Heckman, M.~H\"ubner, and E.~Torres,
  ``{Cornering Relative Symmetry Theories},''
  \href{http://arxiv.org/abs/2408.12600}{{\ttfamily arXiv:2408.12600
  [hep-th]}}.

\bibitem{Bonetti:2024cjk}
F.~Bonetti, M.~Del~Zotto, and R.~Minasian, ``{SymTFTs for Continuous
  non-Abelian Symmetries},'' \href{http://arxiv.org/abs/2402.12347}{{\ttfamily
  arXiv:2402.12347 [hep-th]}}.

\bibitem{Apruzzi:2024htg}
F.~Apruzzi, F.~Bedogna, and N.~Dondi, ``{SymTh for non-finite symmetries},''
  \href{http://arxiv.org/abs/2402.14813}{{\ttfamily arXiv:2402.14813
  [hep-th]}}.

\bibitem{Witten:2009at}
E.~Witten, ``{Geometric Langlands From Six Dimensions},''
  \href{http://arxiv.org/abs/0905.2720}{{\ttfamily arXiv:0905.2720 [hep-th]}}.

\bibitem{Tachikawa:2013hya}
Y.~Tachikawa, ``{On the 6d origin of discrete additional data of 4d gauge
  theories},'' \href{http://dx.doi.org/10.1007/JHEP05(2014)020}{{\em JHEP}
  {\bfseries 05} (2014) 020}, \href{http://arxiv.org/abs/1309.0697}{{\ttfamily
  arXiv:1309.0697 [hep-th]}}.

\bibitem{DelZotto:2015isa}
M.~Del~Zotto, J.~J. Heckman, D.~S. Park, and T.~Rudelius, ``{On the Defect
  Group of a 6D SCFT},''
  \href{http://dx.doi.org/10.1007/s11005-016-0839-5}{{\em Lett. Math. Phys.}
  {\bfseries 106} no.~6, (2016) 765--786},
  \href{http://arxiv.org/abs/1503.04806}{{\ttfamily arXiv:1503.04806
  [hep-th]}}.

\bibitem{Maldacena:2016hyu}
J.~Maldacena and D.~Stanford, ``{Remarks on the Sachdev-Ye-Kitaev model},''
  \href{http://dx.doi.org/10.1103/PhysRevD.94.106002}{{\em Phys. Rev. D}
  {\bfseries 94} no.~10, (2016) 106002},
  \href{http://arxiv.org/abs/1604.07818}{{\ttfamily arXiv:1604.07818
  [hep-th]}}.

\bibitem{Stanford:2019vob}
D.~Stanford and E.~Witten, ``{JT gravity and the ensembles of random matrix
  theory},'' \href{http://dx.doi.org/10.4310/ATMP.2020.v24.n6.a4}{{\em Adv.
  Theor. Math. Phys.} {\bfseries 24} no.~6, (2020) 1475--1680},
  \href{http://arxiv.org/abs/1907.03363}{{\ttfamily arXiv:1907.03363
  [hep-th]}}.

\bibitem{Saad:2019lba}
P.~Saad, S.~H. Shenker, and D.~Stanford, ``{JT gravity as a matrix integral},''
  \href{http://arxiv.org/abs/1903.11115}{{\ttfamily arXiv:1903.11115
  [hep-th]}}.

\bibitem{Balasubramanian:2020lux}
V.~Balasubramanian, J.~J. Heckman, E.~Lipeles, and A.~P. Turner, ``{Statistical
  Coupling Constants from Hidden Sector Entanglement},''
  \href{http://dx.doi.org/10.1103/PhysRevD.103.066024}{{\em Phys. Rev. D}
  {\bfseries 103} no.~6, (2021) 066024},
  \href{http://arxiv.org/abs/2012.09182}{{\ttfamily arXiv:2012.09182
  [hep-th]}}.

\bibitem{Marolf:2020xie}
D.~Marolf and H.~Maxfield, ``{Transcending the ensemble: baby universes,
  spacetime wormholes, and the order and disorder of black hole information},''
  \href{http://dx.doi.org/10.1007/JHEP08(2020)044}{{\em JHEP} {\bfseries 08}
  (2020) 044}, \href{http://arxiv.org/abs/2002.08950}{{\ttfamily
  arXiv:2002.08950 [hep-th]}}.

\bibitem{Heckman:2021vzx}
J.~J. Heckman, A.~P. Turner, and X.~Yu, ``{Disorder Averaging and its UV
  (Dis)Contents},'' \href{http://dx.doi.org/10.1103/PhysRevD.105.086021}{{\em
  Phys. Rev. D} {\bfseries 105} no.~8, (2022) 086021},
  \href{http://arxiv.org/abs/2111.06404}{{\ttfamily arXiv:2111.06404
  [hep-th]}}.

\bibitem{Chandra:2022bqq}
J.~Chandra, S.~Collier, T.~Hartman, and A.~Maloney, ``{Semiclassical 3D gravity
  as an average of large-c CFTs},''
  \href{http://dx.doi.org/10.1007/JHEP12(2022)069}{{\em JHEP} {\bfseries 12}
  (2022) 069}, \href{http://arxiv.org/abs/2203.06511}{{\ttfamily
  arXiv:2203.06511 [hep-th]}}.

\bibitem{Schlenker:2022dyo}
J.-M. Schlenker and E.~Witten, ``{No ensemble averaging below the black hole
  threshold},'' \href{http://dx.doi.org/10.1007/JHEP07(2022)143}{{\em JHEP}
  {\bfseries 07} (2022) 143}, \href{http://arxiv.org/abs/2202.01372}{{\ttfamily
  arXiv:2202.01372 [hep-th]}}.

\bibitem{Benini:2022hzx}
F.~Benini, C.~Copetti, and L.~Di~Pietro, ``{Factorization and global symmetries
  in holography},'' \href{http://dx.doi.org/10.21468/SciPostPhys.14.2.019}{{\em
  SciPost Phys.} {\bfseries 14} no.~2, (2023) 019},
  \href{http://arxiv.org/abs/2203.09537}{{\ttfamily arXiv:2203.09537
  [hep-th]}}.

\bibitem{Wigner:1932eb}
E.~P. Wigner, ``{On the quantum correction for thermodynamic equilibrium},''
  \href{http://dx.doi.org/10.1103/PhysRev.40.749}{{\em Phys. Rev.} {\bfseries
  40} (1932) 749--760}.

\bibitem{WOOTTERS19871}
W.~K. Wootters, ``A wigner-function formulation of finite-state quantum
  mechanics,'' {\em Annals of Physics} {\bfseries 176} no.~1, (1987) 1--21.

\bibitem{Bouzouina1996}
A.~Bouzouina and S.~De~Bi{\`e}vre, ``Equipartition of the eigenfunctions of
  quantized ergodic maps on the torus,''
  \href{http://dx.doi.org/10.1007/BF02104909}{{\em Communications in
  Mathematical Physics} {\bfseries 178} no.~1, (1996) 83--105}.

\bibitem{Bianucci2001DiscreteWF}
P.~Bianucci, C.~Miquel, J.~P. Paz, and M.~Saraceno, ``Discrete wigner functions
  and the phase space representation of quantum computers,'' {\em Physics
  Letters A} {\bfseries 297} (2001) 353--358.

\bibitem{preskill2015lecture}
J.~Preskill, {\em Lecture Notes for Physics 229:Quantum Information and
  Computation}.
\newblock CreateSpace Independent Publishing Platform, 2015.
\newblock \url{https://books.google.se/books?id=MIv8rQEACAAJ}.

\bibitem{CHOI1975285}
M.-D. Choi, ``Completely positive linear maps on complex matrices,''
  \href{http://dx.doi.org/https://doi.org/10.1016/0024-3795(75)90075-0}{{\em
  Linear Algebra and its Applications} {\bfseries 10} no.~3, (1975) 285--290}.
  \url{https://www.sciencedirect.com/science/article/pii/0024379575900750}.

\bibitem{Witten:1988hf}
E.~Witten, ``{Quantum Field Theory and the Jones Polynomial},''
  \href{http://dx.doi.org/10.1007/BF01217730}{{\em Commun. Math. Phys.}
  {\bfseries 121} (1989) 351--399}.

\bibitem{Belov:2005ze}
D.~Belov and G.~W. Moore, ``{Classification of Abelian spin Chern-Simons
  theories},'' \href{http://arxiv.org/abs/hep-th/0505235}{{\ttfamily
  arXiv:hep-th/0505235}}.

\bibitem{Kapustin:2010hk}
A.~Kapustin and N.~Saulina, ``{Topological boundary conditions in abelian
  Chern-Simons theory},''
  \href{http://dx.doi.org/10.1016/j.nuclphysb.2010.12.017}{{\em Nucl. Phys. B}
  {\bfseries 845} (2011) 393--435},
  \href{http://arxiv.org/abs/1008.0654}{{\ttfamily arXiv:1008.0654 [hep-th]}}.

\bibitem{Belov:2006jd}
D.~Belov and G.~W. Moore, ``{Holographic Action for the Self-Dual Field},''
  \href{http://arxiv.org/abs/hep-th/0605038}{{\ttfamily arXiv:hep-th/0605038}}.

\bibitem{Belov:2006xj}
D.~M. Belov and G.~W. Moore, ``{Type II Actions from 11-Dimensional
  Chern-Simons Theories},''
  \href{http://arxiv.org/abs/hep-th/0611020}{{\ttfamily arXiv:hep-th/0611020}}.

\bibitem{Monnier:2013kna}
S.~Monnier, ``{The Global Anomaly of the Self-Dual Field in General
  Backgrounds},'' \href{http://dx.doi.org/10.1007/s00023-015-0423-z}{{\em
  Annales Henri Poincare} {\bfseries 17} no.~5, (2016) 1003--1036},
  \href{http://arxiv.org/abs/1309.6642}{{\ttfamily arXiv:1309.6642 [hep-th]}}.

\bibitem{Apruzzi:2022dlm}
F.~Apruzzi, ``{Higher form symmetries TFT in 6d},''
  \href{http://dx.doi.org/10.1007/JHEP11(2022)050}{{\em JHEP} {\bfseries 11}
  (2022) 050}, \href{http://arxiv.org/abs/2203.10063}{{\ttfamily
  arXiv:2203.10063 [hep-th]}}.

\bibitem{Heckman:2018jxk}
J.~J. Heckman and T.~Rudelius, ``{Top Down Approach to 6D SCFTs},''
  \href{http://dx.doi.org/10.1088/1751-8121/aafc81}{{\em J. Phys. A} {\bfseries
  52} no.~9, (2019) 093001}, \href{http://arxiv.org/abs/1805.06467}{{\ttfamily
  arXiv:1805.06467 [hep-th]}}.

\bibitem{Argyres:2022mnu}
P.~C. Argyres, J.~J. Heckman, K.~Intriligator, and M.~Martone, ``{Snowmass
  White Paper on SCFTs},'' \href{http://arxiv.org/abs/2202.07683}{{\ttfamily
  arXiv:2202.07683 [hep-th]}}.

\bibitem{Lawrie:2023tdz}
C.~Lawrie, X.~Yu, and H.~Y. Zhang, ``{Intermediate defect groups, polarization
  pairs, and noninvertible duality defects},''
  \href{http://dx.doi.org/10.1103/PhysRevD.109.026005}{{\em Phys. Rev. D}
  {\bfseries 109} no.~2, (2024) 026005},
  \href{http://arxiv.org/abs/2306.11783}{{\ttfamily arXiv:2306.11783
  [hep-th]}}.

\bibitem{Heckman:2022suy}
J.~J. Heckman, C.~Lawrie, L.~Lin, H.~Y. Zhang, and G.~Zoccarato, ``{6D SCFTs,
  Center-Flavor Symmetries, and Stiefel-Whitney Compactifications},''
  \href{http://dx.doi.org/10.1103/PhysRevD.106.066003}{{\em Phys. Rev. D}
  {\bfseries 106} no.~6, (2022) 066003},
  \href{http://arxiv.org/abs/2205.03411}{{\ttfamily arXiv:2205.03411
  [hep-th]}}.

\bibitem{Heckman:2022muc}
J.~J. Heckman, M.~H\"ubner, E.~Torres, and H.~Y. Zhang, ``{The Branes Behind
  Generalized Symmetry Operators},''
  \href{http://dx.doi.org/10.1002/prop.202200180}{{\em Fortsch. Phys.}
  {\bfseries 71} no.~1, (2023) 2200180},
  \href{http://arxiv.org/abs/2209.03343}{{\ttfamily arXiv:2209.03343
  [hep-th]}}.

\bibitem{Bonetti:2024etn}
F.~Bonetti, M.~Del~Zotto, and R.~Minasian, ``{SymTFTs and Non-Invertible
  Symmetries of 6d (2,0) SCFTs of Type D from M-theory},''
  \href{http://dx.doi.org/10.1007/JHEP02(2025)156}{{\em JHEP} {\bfseries 02}
  (2025) 156}, \href{http://arxiv.org/abs/2412.07842}{{\ttfamily
  arXiv:2412.07842 [hep-th]}}.

\bibitem{Ma:2024kma}
R.~Ma and A.~Turzillo, ``{Symmetry-Protected Topological Phases of Mixed States
  in the Doubled Space},''
  \href{http://dx.doi.org/10.1103/PRXQuantum.6.010348}{{\em PRX Quantum}
  {\bfseries 6} no.~1, (2025) 010348},
  \href{http://arxiv.org/abs/2403.13280}{{\ttfamily arXiv:2403.13280
  [quant-ph]}}.

\bibitem{HUDSON1974249}
R.~Hudson, ``When is the wigner quasi-probability density non-negative?,''
  \href{http://dx.doi.org/https://doi.org/10.1016/0034-4877(74)90007-X}{{\em
  Reports on Mathematical Physics} {\bfseries 6} no.~2, (1974) 249--252}.
  \url{https://www.sciencedirect.com/science/article/pii/003448777490007X}.

\bibitem{Gross:2006wkl}
D.~Gross, ``{Hudson{\textquoteright}s theorem for finite-dimensional quantum
  systems},'' \href{http://dx.doi.org/10.1063/1.2393152}{{\em J. Math. Phys.}
  {\bfseries 47} no.~12, (2006) 122107},
  \href{http://arxiv.org/abs/quant-ph/0602001}{{\ttfamily
  arXiv:quant-ph/0602001}}.

\bibitem{PhysRevA.44.R2775}
C.~T. Lee, ``Measure of the nonclassicality of nonclassical states,''
  \href{http://dx.doi.org/10.1103/PhysRevA.44.R2775}{{\em Phys. Rev. A}
  {\bfseries 44} (Sep, 1991) R2775--R2778}.
  \url{https://link.aps.org/doi/10.1103/PhysRevA.44.R2775}.

\bibitem{Kitaev:1997wr}
A.~Y. Kitaev, ``{Fault tolerant quantum computation by anyons},''
  \href{http://dx.doi.org/10.1016/S0003-4916(02)00018-0}{{\em Annals Phys.}
  {\bfseries 303} (2003) 2--30},
  \href{http://arxiv.org/abs/quant-ph/9707021}{{\ttfamily
  arXiv:quant-ph/9707021}}.

\bibitem{PhysRevA.71.022316}
S.~Bravyi and A.~Kitaev, ``Universal quantum computation with ideal clifford
  gates and noisy ancillas,''
  \href{http://dx.doi.org/10.1103/PhysRevA.71.022316}{{\em Phys. Rev. A}
  {\bfseries 71} (Feb, 2005) 022316}.
  \url{https://link.aps.org/doi/10.1103/PhysRevA.71.022316}.

\bibitem{Hubner:2022kxr}
M.~Hubner, D.~R. Morrison, S.~Schafer-Nameki, and Y.-N. Wang, ``{Generalized
  Symmetries in F-theory and the Topology of Elliptic Fibrations},''
  \href{http://dx.doi.org/10.21468/SciPostPhys.13.2.030}{{\em SciPost Phys.}
  {\bfseries 13} no.~2, (2022) 030},
  \href{http://arxiv.org/abs/2203.10022}{{\ttfamily arXiv:2203.10022
  [hep-th]}}.

\bibitem{Cvetic:2022imb}
M.~Cveti\v{c}, J.~J. Heckman, M.~H\"ubner, and E.~Torres, ``{0-Form, 1-Form and
  2-Group Symmetries via Cutting and Gluing of Orbifolds},''
  \href{http://arxiv.org/abs/2203.10102}{{\ttfamily arXiv:2203.10102
  [hep-th]}}.

\bibitem{GarciaEtxebarria:2024fuk}
I.~Garcia~Etxebarria and S.~S. Hosseini, ``{Some Aspects of Symmetry
  Descent},'' \href{http://dx.doi.org/10.1007/JHEP12(2024)223}{{\em JHEP}
  {\bfseries 12} (2025) 223}, \href{http://arxiv.org/abs/2404.16028}{{\ttfamily
  arXiv:2404.16028 [hep-th]}}.

\bibitem{Gukov:2020btk}
S.~Gukov, P.-S. Hsin, and D.~Pei, ``{Generalized Global Symmetries of $T[M]$
  Theories. I},'' \href{http://dx.doi.org/10.1007/JHEP04(2021)232}{{\em JHEP}
  {\bfseries 04} (2021) 232}, \href{http://arxiv.org/abs/2010.15890}{{\ttfamily
  arXiv:2010.15890 [hep-th]}}.

\bibitem{Apruzzi:2016nfr}
F.~Apruzzi, F.~Hassler, J.~J. Heckman, and I.~V. Melnikov, ``{From 6D SCFTs to
  Dynamic GLSMs},'' \href{http://dx.doi.org/10.1103/PhysRevD.96.066015}{{\em
  Phys. Rev. D} {\bfseries 96} no.~6, (2017) 066015},
  \href{http://arxiv.org/abs/1610.00718}{{\ttfamily arXiv:1610.00718
  [hep-th]}}.

\bibitem{Cvetic:2023pgm}
M.~Cveti\v{c}, J.~J. Heckman, M.~H\"ubner, and E.~Torres, ``{Generalized
  Symmetries, Gravity, and the Swampland},''
  \href{http://dx.doi.org/10.1103/PhysRevD.109.026012}{{\em Phys. Rev. D}
  {\bfseries 109} no.~2, (2024) 026012},
  \href{http://arxiv.org/abs/2307.13027}{{\ttfamily arXiv:2307.13027
  [hep-th]}}.

\bibitem{Witten:1999xp}
E.~Witten and S.-T. Yau, ``{Connectedness of the boundary in the AdS / CFT
  correspondence},'' \href{http://dx.doi.org/10.4310/ATMP.1999.v3.n6.a1}{{\em
  Adv. Theor. Math. Phys.} {\bfseries 3} (1999) 1635--1655},
  \href{http://arxiv.org/abs/hep-th/9910245}{{\ttfamily arXiv:hep-th/9910245}}.

\bibitem{Maldacena:2004rf}
J.~M. Maldacena and L.~Maoz, ``{Wormholes in AdS},''
  \href{http://dx.doi.org/10.1088/1126-6708/2004/02/053}{{\em JHEP} {\bfseries
  02} (2004) 053}, \href{http://arxiv.org/abs/hep-th/0401024}{{\ttfamily
  arXiv:hep-th/0401024}}.

\bibitem{Aharony:1998qu}
O.~Aharony and E.~Witten, ``{Anti-de Sitter space and the center of the gauge
  group},'' \href{http://dx.doi.org/10.1088/1126-6708/1998/11/018}{{\em JHEP}
  {\bfseries 11} (1998) 018},
  \href{http://arxiv.org/abs/hep-th/9807205}{{\ttfamily arXiv:hep-th/9807205}}.

\bibitem{Maldacena:2001ss}
J.~M. Maldacena, G.~W. Moore, and N.~Seiberg, ``{D-brane charges in five-brane
  backgrounds},'' \href{http://dx.doi.org/10.1088/1126-6708/2001/10/005}{{\em
  JHEP} {\bfseries 10} (2001) 005},
  \href{http://arxiv.org/abs/hep-th/0108152}{{\ttfamily arXiv:hep-th/0108152}}.

\bibitem{Heckman:2022xgu}
J.~J. Heckman, M.~Hubner, E.~Torres, X.~Yu, and H.~Y. Zhang, ``{Top Down
  Approach to Topological Duality Defects},''
  \href{http://dx.doi.org/10.1103/PhysRevD.108.046015}{{\em Phys. Rev. D}
  {\bfseries 108} no.~4, (2023) 046015},
  \href{http://arxiv.org/abs/2212.09743}{{\ttfamily arXiv:2212.09743
  [hep-th]}}.

\bibitem{Israel:1976ur}
W.~Israel, ``{Thermo-field dynamics of black holes},''
  \href{http://dx.doi.org/10.1016/0375-9601(76)90178-X}{{\em Phys. Lett. A}
  {\bfseries 57} (1976) 107--110}.

\bibitem{Maldacena:2001kr}
J.~M. Maldacena, ``{Eternal black holes in Anti-de-Sitter},''
  \href{http://dx.doi.org/10.1088/1126-6708/2003/04/021}{{\em JHEP} {\bfseries
  04} (2003) 021}, \href{http://arxiv.org/abs/hep-th/0106112}{{\ttfamily
  arXiv:hep-th/0106112}}.

\bibitem{Maldacena:2013xja}
J.~Maldacena and L.~Susskind, ``{Cool horizons for entangled black holes},''
  \href{http://dx.doi.org/10.1002/prop.201300020}{{\em Fortsch. Phys.}
  {\bfseries 61} (2013) 781--811},
  \href{http://arxiv.org/abs/1306.0533}{{\ttfamily arXiv:1306.0533 [hep-th]}}.

\end{thebibliography}\endgroup

\end{document}